\documentclass{article}

\usepackage{PRIMEarxiv}

\usepackage[utf8]{inputenc} 
\usepackage[T1]{fontenc}    
\usepackage{hyperref}       
\usepackage{url}            
\usepackage{booktabs}       
\usepackage{amsfonts}       
\usepackage{nicefrac}       
\usepackage{microtype}      
\usepackage{lipsum}
\usepackage{fancyhdr}       
\usepackage{graphicx}       
\graphicspath{{media/}}     
\usepackage{natbib}
\usepackage{amsmath}
\usepackage{amssymb}
\usepackage{mathrsfs}
\usepackage{xcolor}
\title{Elucidating three-dimensional coherent structures in a multi-stream jet}

\author{
  Mitesh Thakor\thanks{\texttt{corresponding author, currently at The Ohio State University}} \\
  Department of Mechanical and Aerospace Engineering\\
  Syracuse University\\
  Syracuse, NY 13210 \\
  \texttt{thakor.6@osu.edu} \\
   \And
  Datta V. Gaitonde \\ 
  Department of Mechanical and Aerospace Engineering\\
  The Ohio State University\\
  Columbus, OH 43210\\
  \texttt{gaitonde.3@osu.edu} \\
   \And
  Yiyang Sun \\ 
  Department of Mechanical and Aerospace Engineering\\
  Syracuse University\\
  Syracuse, NY 13210 \\
  \texttt{ysun58@syr.edu} \\
}

\begin{document}
\maketitle

\begin{abstract}

Nominal two-dimensional (2D) shear layers have been studied extensively, and their principal dynamics are well understood. 
In practical configurations, however, the behavior of such shear layers is affected by proximal surfaces.
In this study, we investigate three-dimensional (3D) coherent structures developing downstream of a relatively thick splitter plate in a realistic nozzle featuring sidewalls, an upper boundary formed by a single expansion ramp, and a lower boundary defined by a protruding deck.
As a result, in addition to the \textit{primary} splitter plate shear layer (SPSL) arising from mixing between the core and bypass streams, the flow contains \textit{upper} (USL) and \textit{lower} (LSL) shear layers with the ambient.
Large-eddy simulation data are analyzed to characterize the unsteady flow dynamics, while the mean flow provides insight into the underlying amplification mechanisms. 
Spectral proper orthogonal decomposition reveals a clear separation of broadband and tonal dynamics across frequency bands.
The broadband low-frequency modes are highly 3D and originate in the USL and LSL.
In contrast, tonal high-frequency content is associated with a 2D instability in the SPSL.
Both the broadband and tonal signatures also appear in the nonlinear energy transfer mechanisms.
Triglobal resolvent analysis further clarifies the amplification mechanisms within the USL and LSL.
Low-frequency response modes are excited by forcing localized near the nozzle geometry and are governed by 3D Kelvin–Helmholtz dynamics. 
The low-frequency streamwise vortices generated at the nozzle corners drive the axis-switching behavior characteristic of rectangular jets.
Wavemaker analysis further demonstrates that these corner vortices are part of self-sustaining low-frequency dynamics.

\end{abstract}


\section{Introduction} \label{sec:intro}

Jet flows have long been a central subject in turbulence research due to their importance in propulsion systems and their relatively canonical nature.
The early evolution of turbulent jets is now widely understood to be governed by large-scale structures that emerge at the jet boundaries \citep{crow1971, list1982turbulent}.
These structures, produced by the rapid instability and roll-up of the initial shear layers near the nozzle exit, play a decisive role in ambient-fluid entrainment and in the generation of jet noise \citep{becker1968vortex, jordan2013}.
In axisymmetric jets, the rapid destabilization of the initial shear layers by Kelvin-Helmholtz instabilities gives rise to organized vortex rings \citep{suzuki2006instability, schmidt2018}.
Such large-scale structures are therefore fundamental to both the near-field dynamics and the far-field acoustic signature of turbulent jets.

Over the past several decades, jet engines have advanced considerably, moving from relatively simple axisymmetric nozzles to more intricate non-axisymmetric geometries \citep{capone1979}.
Rectangular nozzles, in particular, have received extensive attention because of the ease of airframe-propulsion integration.
The resulting rectangular jets exhibit fundamentally non-axisymmetric flow physics \citep{kantola1979noise, krothapalli1981, krothapalli1986role, tam1997screech} with inherently more complex flow than that of axisymmetric jets due to the three-dimensional (3D) effects introduced by the nozzle corners.
A defining feature is axis-switching, which occurs from the differential growth rates of the shear layers along the major and minor axes of the jet.
Hence, the axes exchange happens as the jet evolves downstream and the plume gradually transitions toward an axisymmetric state \citep{sfeir1979investigation, krothapalli1981, ho1987, hussain1989}.
The primary mechanisms governing axis-switching are the azimuthal and streamwise vorticity dynamics.
The counter-rotating corner vortices can either promote or suppress axis-switching depending on whether they entrain or reject fluid along the corners \citep{zaman1996}.

\begin{figure}
     \centering
         \includegraphics[width=1\textwidth]{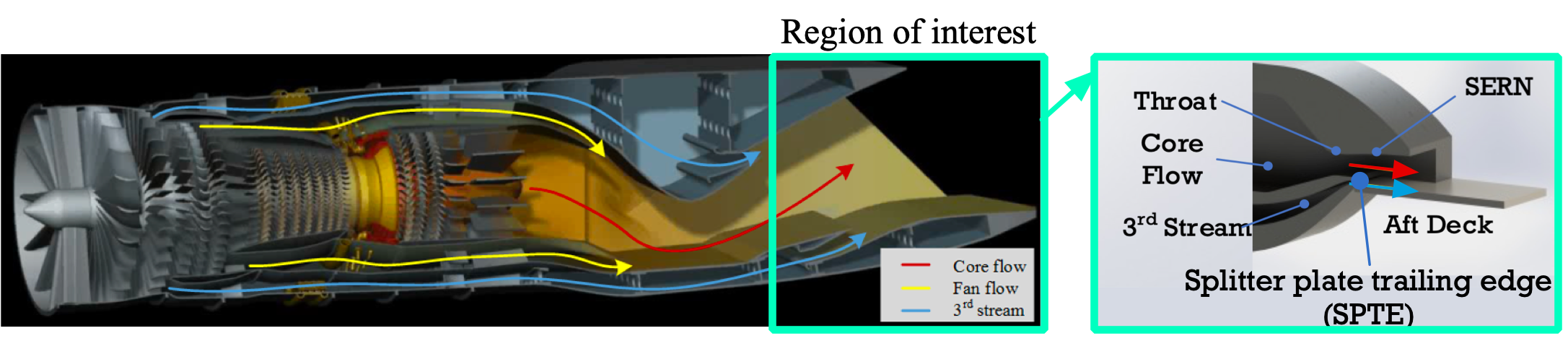}
    \caption{Three-stream turbofan configuration based on the Air Force Research Laboratory (AFRL) design \citep{simmons2009}. The figure shows an isometric cross-section of the nozzle layout, with the core stream (Mach 1.6) marked in red and the third bypass stream (Mach 1) marked in blue. SERN: Single-sided Expansion Ramp Nozzle.}
    \label{fig:engine}
\end{figure}

In this study, the configuration of particular interest is the non-axisymmetric, airframe-integrated, variable-cycle engine, shown in figure~\ref{fig:engine} \citep{simmons2009}.
The focus in this work is on the nozzle region, highlighted in the inset.
The rectangular nozzle contains multiple streams; the merged core and fan streams form the primary (or collectively designated as `core') jet.
An additional third bypass stream, referred to as the bypass stream, is introduced in the expansion part of the nozzle; this stream mixes downstream of the relatively thick splitter plate trailing edge (SPTE) to form the primary shear layer.
The bypass stream offers thermodynamic advantages that can improve overall efficiency and reduce jet noise \citep{papamoschou2001}.
On the upper side, the core stream is bounded by the Single Expansion Ramp Nozzle (SERN), while on the lower side, the bypass stream emerges over an aft deck.  
This nozzle configuration has been studied extensively, both numerically \citep{stack2018, doshi2022} and experimentally \citep{berry2016, magstadt2017}.
The measured far-field acoustic spectra exhibit a low-frequency broadband and a high-frequency tonal behavior.
The latter component has been especially examined, and may be traced to shedding from the SPTE as the core and bypass streams mix under the influence of a vortex-shedding instability \citep{doshi2022, thakor2024, thakor2025}.

The understanding derived from these studies has recently motivated flow control efforts to modulate the vortex shedding and thereby reduce high-frequency noise generation \citep{kelly2024schlieren, yeung2024high, qualters2025experimental}.
Both numerical and experimental efforts have investigated steady and unsteady blowing through the splitter plate or aft-deck.
While these studies have focused primarily on the mechanisms underlying high-frequency tonal behavior, the low-frequency dynamics of this jet remain relatively unexplored, despite their substantial contribution to overall noise levels \citep{magstadt2017, kelly2024schlieren}.
In this work, we focus specifically on these low-frequency dynamics and aim to uncover the physical mechanisms responsible for them, including their interactions with the high-frequency phenomena, using both data-driven as well as operator-based methods.

Coherent structures in turbulent flows can be extracted directly from high-fidelity datasets or modeled through the governing equations, applied to the mean flow as a basic state \citep{rowley2017, taira2017}.
For statistically stationary flows, spectral proper orthogonal decomposition (SPOD) of the scale-resolved data is a natural approach for identifying and analyzing energy-ranked coherent structures that oscillate at a single frequency \citep{lumley1967, glauser1987, towne2018}.
In parallel, linear-operator-based techniques have proven highly effective for predicting and modeling coherent structures using only a steady flowfield \citep{schmid2002}.
Resolvent analysis, in particular, characterizes how harmonic forcing at a given frequency is amplified by the linearized Navier-Stokes operator relative to a given base state.
For turbulent flows, the time-averaged flow serves as the basic state, and the nonlinear terms are interpreted as intrinsic forcing \citep{mckeon2010}.
Theoretically, the resolvent response modes are equivalent to SPOD modes under the assumption of white-noise nonlinear forcing \citep{towne2018}; however, the correlated and structured nature of nonlinear forcing in turbulent flows can limit direct comparisons between resolvent predictions and turbulent statistics.
Resolvent analysis has been applied across a broad range of flows to identify key amplification mechanisms, including open cavities \citep{sun2020}, airfoils \citep{symon2018, ribeiro2023}, axisymmetric jets \citep{garnaud2013, jeun2016, schmidt2018}, and shear layers \citep{doshi2022, thakor2024}.
In this study, we extract coherent structures using SPOD and resolvent analyses; given the emphasis on the finite-span aspect of the problem, both approaches are invoked in their full 3D forms without invoking spatial homogeneity.

The paper is organized as follows. \S\ref{sec:baseline} describes the large-eddy simulation foundational dataset of the multi-stream rectangular jet. \S\ref{sec:datadriven} introduces the data-driven framework, beginning with the methodology and results of spectral proper orthogonal decomposition in \S\ref{sec:spod}. \S\ref{sec:bmd} examines the nonlinear interactions across the various shear layers using bispectral mode decomposition. \S\ref{sec:resl} presents the formulation and results of the triglobal resolvent analysis. Finally, the conclusions are discussed in \S\ref{sec:concl}.

\section{Large-eddy simulation of rectangular jet flow} \label{sec:baseline}

Figure~\ref{fig:engine} shows the nozzle configuration examined in this study. For reference, we summarize the flow conditions and key geometric features here.
As noted earlier, although designated a three-stream engine, the core and fan streams are fully mixed and, in this work, are treated as a single core stream. The bypass (third) stream is injected in the diverging section of the nozzle.
It emerges between the splitter plate and the aft-deck, which represents the aircraft frame, and extends two hydraulic diameters ($D_h$) from the nozzle exit.
The rectangular nozzle cross-section has an aspect ratio of $2.7$ at the exit. The hydraulic diameter is approximately $44 mm$, and the splitter-plate width ($w_{sp}$) is about $1.85 D_h$.
The operation conditions are specified in terms of the nozzle pressure ratio (NPR) and nozzle temperature ratio (NTR), which are the ratios of stagnation pressure and stagnation temperature to their ambient values.
The core and bypass streams operate at nozzle pressure ratios of $\text{NPR}_{\text{core}} = 4.25$ and $\text{NPR}_{\text{bypass}} = 1.89$, corresponding to Mach numbers of $1.6$ and $1.0$, respectively, based on isentropic relations.
Both streams have a nozzle temperature ratio of $\text{NTR}_{\text{core}} = \text{NTR}_{\text{bypass}} = 1$.  Further information on the nozzle geometry and design methodology may be found in \citep{berry2016} and \citep{magstadt2017}.

The benchmark Large-eddy simulations (LES) were carried out as described in detail in \citep{stack2018}, including extensive mesh resolution studies.
The simulations were extensively validated against companion experiment data \citep{berry2016, magstadt2017}, including pressure spectra, schlieren, and Particle Image Velocimetry (PIV) measurements.
The splitter-plate width ($w_{sp}$), which corresponds to the major axis of the nozzle, is used to normalize the coordinate system.
The jet Reynolds number is defined as $Re_j= \rho_j \overline{u}_j D_h/\mu_j = 2.7 \times 10^6$, based on the nozzle exit density ($\rho_j$), mean exit velocity ($\overline{u}_j$), and viscosity ($\mu_j$).
These properties are obtained from isentropic relations at the design Mach number.
Here, the subscript $j$ denotes jet-exit quantities.
Frequencies are reported using the Strouhal number, $St_{D_h} = \omega D_h/2\pi \overline{u}_j$, where $\omega$ is the angular frequency.
For simplicity, we omit the $D_h$ subscript and refer to the Strouhal number simply as $St$.
Velocity components in the streamwise, transverse, and spanwise directions are denoted by $u$, $v$, and $w$, corresponding to the $x$, $y$, and $z$ directions.
Pressure and temperature are represented by $P$ and $T$, respectively.
\begin{figure}
     \centering
         \includegraphics[width=0.9\textwidth]{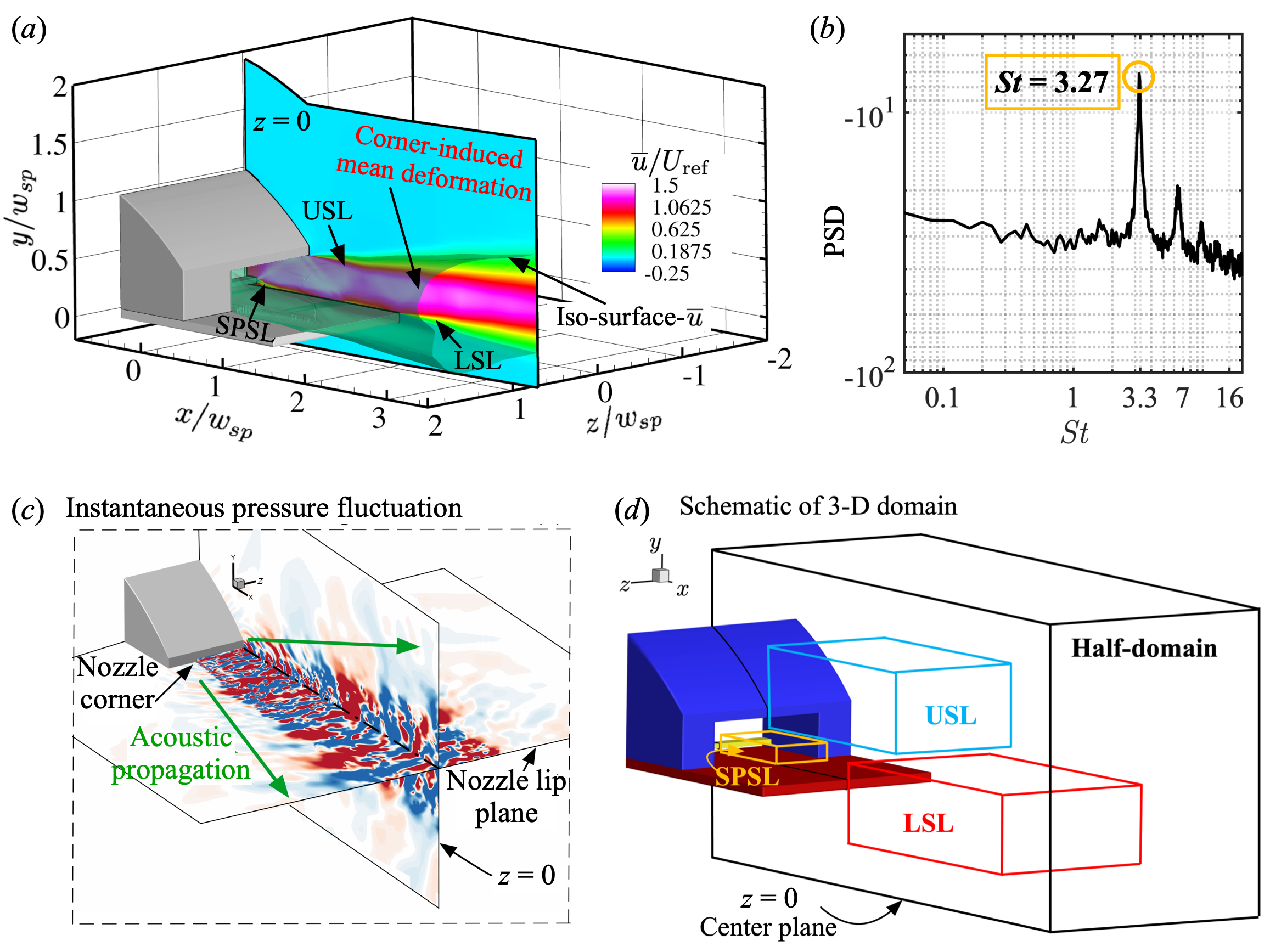}
         \caption{(a) Normalized time-averaged streamwise velocity ($\overline{u}/U_{\text{ref}}$) shows three main shear layers, splitter plate shear layer (SPSL), upper shear layer (USL), and lower shear layer (LSL), in the flowfield for nozzle center plane ($z=0$). Iso-surface at level of $\overline{u}/U_{\text{ref}} = 0.4$. (b) Power spectrum density (PSD) of $u$ velocity at the probe located in the SPSL region. (c) Instantaneous pressure fluctuation ($P'/P_{\text{ref}}$) \textcolor{blue}{\rule{1.5ex}{1.5ex}}\textcolor{red}{\rule{1.5ex}{1.5ex}} (blue, red) $\in$ [-0.1, 0.1].  (d) Schematic of 3D domain utilized to compute spectral analysis (not to scale).}
    \label{fig:u_mean}
\end{figure}

Figure~\ref{fig:u_mean}(a) shows the normalized, time-averaged streamwise velocity ($\overline{u}/U_{\text{ref}}$) at the nozzle center plane ($z=0$). 
The velocities are normalized using the ambient speed of sound, $U_{\text{ref}}$.
The three main shear layers of interest are marked in the mean flow: the splitter-plate shear layer (SPSL), which forms as the core and bypass streams mix; the upper shear layer (USL), created by mixing between the jet and the ambient flow on the upper side of the nozzle; and the lower shear layer (LSL), produced by the corresponding mixing downstream of the aft-deck on the lower side.
An oblique shock forms at the SPTE and generates a shock train as it reflects off the SERN, impinges on the aft-deck, and reflects again.
The iso-surface of time-averaged streamwise velocity highlights the 3D effects generated by the nozzle corner in the mean flow manifested as kinks; the one associated with the upper right corner is marked in the figure~\ref{fig:u_mean}(a).
We will examine these effects in the following \S\ref{sec:corner}.

Figure~\ref{fig:u_mean}(b) presents the power spectral density of the streamwise velocity at a probe located in the SPSL. The probe captures a dominant vortex-shedding frequency at $St = 3.27$, which corresponds to roughly $34 kHz$.
Experiments capture the same peaks in the far-field spectra \citep{berry2016, magstadt2017}, indicating the global dominance of this feature. Some of the main physical mechanisms and manifestations associated with these have been examined in our companion studies \citep{thakor2024, thakor2025}.
The instantaneous pressure fluctuations, shown in figure~\ref{fig:u_mean}(c) reveal multiscale, 3D coherent structures near the nozzle corner, as well as far-field acoustic radiation along both the major and minor axis planes.
The focus of the present work is on the structure and associated spectral dynamics underlying large-scale structures that originate from the nozzle.
 
To isolate the physical mechanisms associated with each shear layer, these shear layers are treated as spatially separated and analyzed individually, as illustrated in figure~\ref{fig:u_mean}(d).
Due to the symmetrical nature of the nozzle geometry, only half-domain data is considered with the nozzle center plane ($z=0$) as the symmetric plane. 
The following sections present a detailed analysis of the dynamics in each shear layer region and their collective influence on the overall jet dynamics.

\section{Modal analysis of LES data} \label{sec:datadriven}

Modal decomposition methods reveal frequency-resolved spatial support of coherent structures in turbulent flows \citep{rowley2017, taira2017}.
Spectral proper orthogonal decomposition can identify energy-ranked modes that oscillate at a single frequency in statistically stationary flows \citep{glauser1987}. Its utility has greatly increased due to recent efforts that have clearly delineated its properties \citep{towne2018} and the availability of detailed, spatio-temporally resolved flow data.
While SPOD is effective at capturing dominant dynamics, it does not reveal the nonlinear interactions that occur in the flowfield. To address this limitation, we employ bispectral mode decomposition (BMD), which incorporates triadic interactions arising from the quadratic nonlinearity of the Navier--Stokes equations \citep{schmidt2020}.

In the following sections, we apply SPOD and BMD, respectively, to analyze the 3D energy-ranked coherent structures and their interactions with one another in the multi-stream jet flow.
Together, the findings from applying these approaches provide complementary insight into the linear and nonlinear mechanisms that govern flow dynamics.

\subsection{Spectral proper orthogonal decomposition} \label{sec:spod}
SPOD computes a set of orthogonal spatio-temporally correlated modes that optimally capture turbulent flow statistics \citep{towne2018}.
For completeness, a summary of the approach follows.
The LES datasets are first segmented into blocks $\boldsymbol{Q} = [\boldsymbol{q}^{(1)}, \boldsymbol{q}^{(2)}, ..., \boldsymbol{q}^{(n_\text{freq})}]$, where each block or segment contains $n_\text{freq}$ snapshots with an overlap.
A periodic Hanning window is used over each block to prevent spectral leakage. After applying the temporal Fourier transform, we obtain $\boldsymbol{\hat{Q}}_{\omega_k}^{(l)} = [\boldsymbol{\hat{q}}_{ \omega_1}^{(l)}, \boldsymbol{\hat{q}}_{\omega_2}^{(l)}, ..., \boldsymbol{\hat{q}}_{\omega_{n_\text{freq}}}^{(l)} ] $, where $\boldsymbol{\hat{q}}_{\omega_k}^{(l)}$ is the $l$th block Fourier realization of the transformation at the $k$th frequency ($\omega_k$). The cross-frequency density tensor at the given frequency is computed by $\boldsymbol{\hat{S}}_{\omega_k} = \boldsymbol{\hat{Q}}_{\omega_k} \boldsymbol{\hat{Q}}_{\omega_k}^*$, where $(\cdot)^*$ denotes the Hermitian transpose. The SPOD eigenvalue problem can be solved as \citep{lumley1967}
\begin{equation}
\boldsymbol{\hat{S}}_{\omega_k}  \boldsymbol{W} \boldsymbol{\Psi}_{\omega_k} = \boldsymbol{\Psi}_{\omega_k} \boldsymbol{\Lambda}_{\omega_k},
\label{eq:csd}
\end{equation}
where the SPOD modes are presented by each column of $\boldsymbol{\Psi}_{\omega_k}$, ranked by the eigenvalues $\boldsymbol{\Lambda}_{\omega_k} = \text{diag}(\lambda_1, \lambda_2, ..., \lambda_N)$. The SPOD modes are orthonormal ($\boldsymbol{\Psi}_{\omega_k}^* \boldsymbol{W} \boldsymbol{\Psi}_{\omega_k} = \boldsymbol{I}$) and optimal in terms of modal energy with respect to the norm $\boldsymbol{W}$. In this study, we consider the compressible energy norm \citep{chu1965} defined by
\begin{equation}
E = \int_V \left[  \frac{R \overline{T}}{\overline{\rho}} \rho'^2 +  \overline{\rho} u_i' u_i' + \frac{R \overline{\rho}}{(\gamma-1)\overline{T}}  T'^2 \right]  \,dV.
\label{eq:E_chu}
\end{equation}

LES snapshots are interpolated onto a uniform grid with approximately $4.1 \times 10^6$ grid points and state variables $[\rho', u', v', w', T']$.
As noted earlier, to reduce the computational expenses associated with this analysis, configuration symmetry is exploited by considering only half-domain data from the center plane in the spanwise direction, and the different regions of interest are treated separately, taking care to account for the fully 3D nature of the solution that incurs features such as corner vortices and axis switching.
Based on the convergence of the coherent structures and frequency resolution, $6{,}845$ snapshots, at time intervals of $\triangle t U_{\text{ref}}/{w_{sp}} \approx 0.016$ with the convective time length $t_c U_{\text{ref}}/ w_{sp} \approx 110$, are segmented into $23$ blocks, each block containing $1{,}024$ snapshots with $75\%$ overlap.

\begin{figure}
     \centering
         \includegraphics[width=1\textwidth]{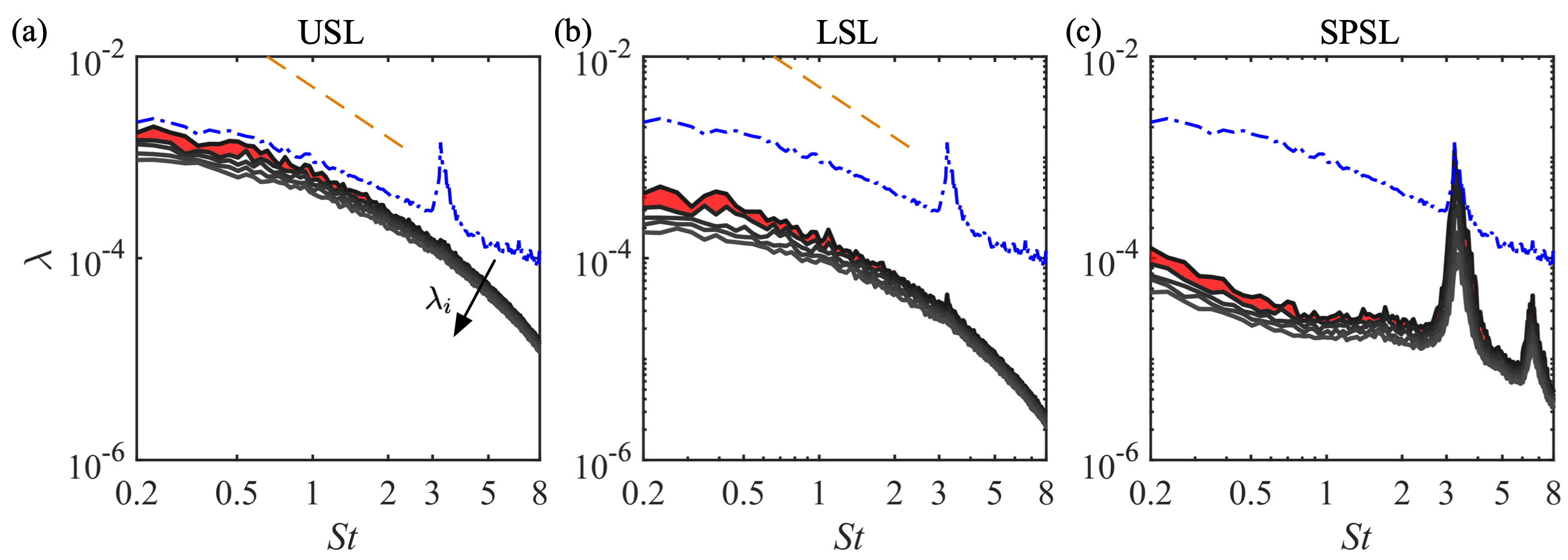}
    \caption{The leading five SPOD eigenvalue spectra for (a) upper shear layer (USL), (b) lower shear layer (LSL), and (c) splitter plate shear layer (SPSL). The blue dot-dashed line denotes the leading eigenvalue computed for the entire half-domain in each plot. The dashed orange line indicates the $-5/3$ power law of the energy spectrum. At each frequency, decreasing eigenvalues are shown in lighter shades, i.e., $\lambda_1 \geq \lambda_2 \dots \geq \lambda_5$.}
    \label{fig:spod_spectra}
\end{figure}

Figure~\ref{fig:spod_spectra} shows the spectrum of SPOD eigenvalues based on Chu's energy in the different shear layers: USL, LSL, and SPSL. Each plot shows the leading five eigenvalues, with the red-shaded area representing the separation between the leading first and second eigenvalues. The blue dot-dashed line denotes the leading eigenvalue ($\lambda_1$) computed for the entire half-domain.
Both the USL (figure~\ref{fig:spod_spectra}(a)) and LSL (figure~\ref{fig:spod_spectra}(b)) regions display broadband behavior.
In these regions, the energy content decreases with increasing frequency and exhibits an approximate $-5/3$ power-law decay.
At the lower frequencies, $St \leq 1$, the first and second eigenvalues are marginally separated, as highlighted by a red-shaded area. No apparent low-rank behavior (i.e., when $\lambda_1 \gg \lambda_2$) is observed for any shear layer regions, likely due to the 3D nature of the flow, which distributes energy among multiple modes.
On the other hand, the SPSL region (see figure~\ref{fig:spod_spectra}(c)) shows a pronounced spectral peak at $St = 3.225$ ($f \approx 34 \text{ kHz}$ ), indicative of a tonal nature.
The leading eigenvalue of the entire half-domain (shown by the blue dot-dashed line) is, by construction, combining the effects of all three shear layers.
A key highlight from the SPOD spectrum when comparing features from each shear layer region to the entire jet flow result is that USL and LSL are primarily governed by the broadband behavior of the low frequencies ($St \leq 1.5$), whereas the SPSL contributes to the tonal nature at the high frequencies ($St \geq 3$) of the overall spectra.

These SPOD spectral trends are consistent with our previous work focusing on the center plane region \citep{thakor2025}, although the current analysis includes additional complexity from the corner flow due to three
dimensionality, which were not accounted for earlier, but are further elaborated on later.

\begin{figure}
     \centering
         \includegraphics[width=1\textwidth]{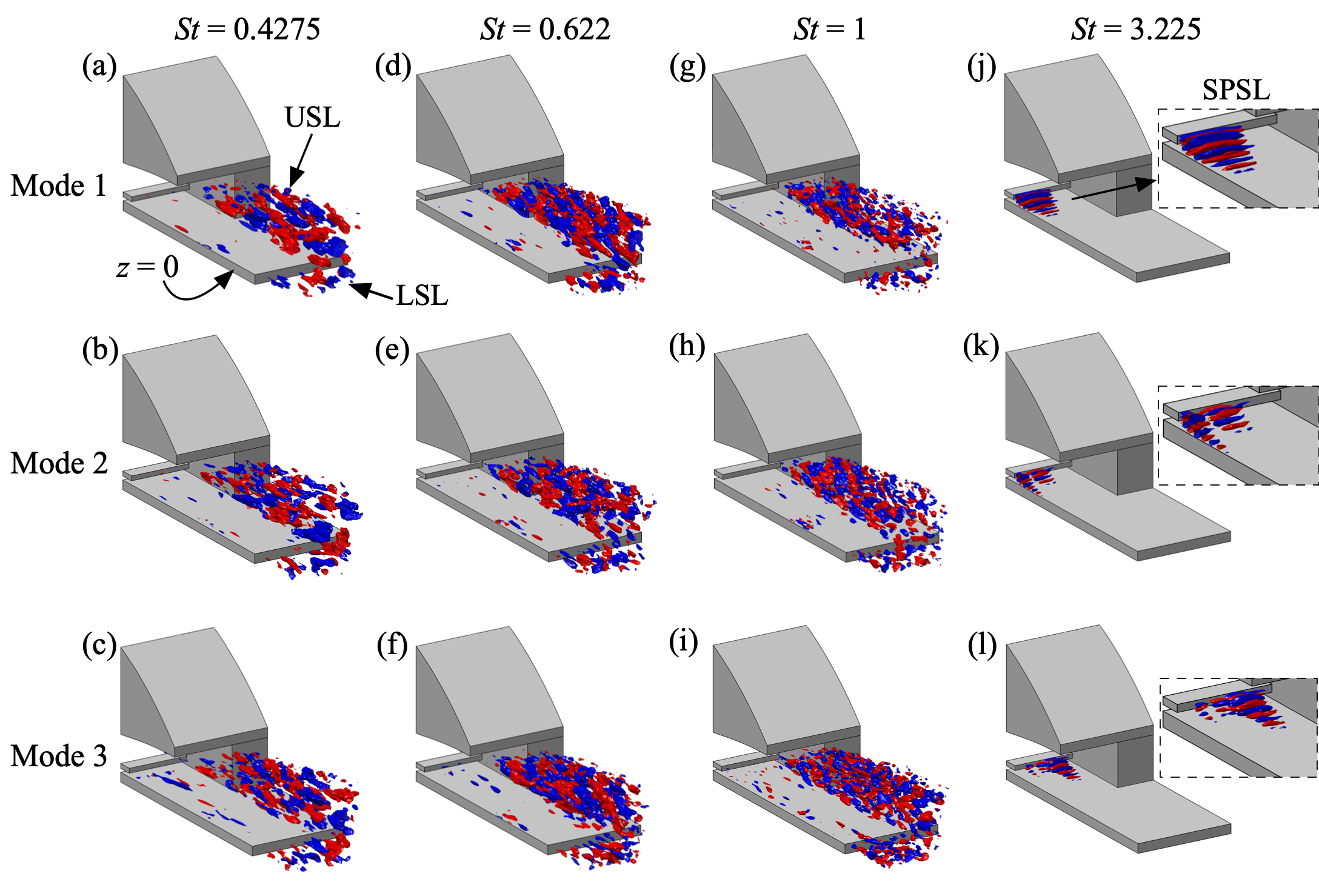}
    \caption{The leading three SPOD modes for the half-domain case. (a-c) $St = 0.4275$, (d-f) $0.622$, (g-i) $1$, and (j-l) 3.225. Iso-surface of the real component of $u$-velocity is shown. \textcolor{blue}{\rule{1.5ex}{1.5ex}}\textcolor{red}{\rule{1.5ex}{1.5ex}} (blue, red) = [-0.3, 0.3].}
    \label{fig:spod_modes}
\end{figure}

Figure~\ref{fig:spod_modes} shows the leading three SPOD modes at representative frequencies for the half-domain case.
For brevity, SPOD modes corresponding to the individual shear layer regions are omitted, as their spatial structures closely resemble those observed in the half-domain within the respective regions.
At low frequencies ($St \leq 1$), the USL and LSL exhibit highly 3D structures across the leading three modes (figures~\ref{fig:spod_modes}(a–i)).
The spatial structures of the modes appear rather incoherent, consistent with the high-rank behavior in the SPOD spectra and the overall 3D nature of these shear layers.
The wavelength of the spatial structures decreases as the frequency increases from $St=0.4275$ to $1$.

At the dominant tonal frequency ($St = 3.225$), the leading SPOD mode displays prominent 2D Kelvin-Helmholtz (KH) roller structures in the SPSL as shown in figure~\ref{fig:spod_modes}(j).
The subsequent modes reveal spanwise variations, highlighting the 3D nature of the SPSL (figures~\ref{fig:spod_modes}(k-l)).
Overall, the 3D SPOD modes confirm the dominance of the primary 2D KH instability originating from the mixing of the main and bypass streams near the SPTE.
These modal structures emphasize the finding from the SPOD spectra, that low-frequency dynamics ($St \leq 1.5$) primarily govern the USL and LSL, while high-frequency dynamics ($St \geq 3$) are predominantly associated with the SPSL.

\subsection{Nonlinear interactions in the multiple shear layers} \label{sec:bmd}

Nonlinear interactions between coherent structures are a fundamental feature of turbulent shear flows and play a key role in redistributing energy across scales and frequencies.
Bispectral mode decomposition isolates triadic interactions in which two frequencies nonlinearly couple to generate a third, offering direct insight into mechanisms such as vortex pairing, subharmonic growth, and interscale energy transfer. 
Recent studies on axisymmetric jet flows~\citep{schmidt2020, nekkanti2025} show that triadic interactions involving Kelvin–Helmholtz wavepackets contribute to the formation of larger-scale structures and streamwise vortices. These findings motivate the use of BMD here to elucidate nonlinear coupling across the multiple shear layers in the present study.

Aspects of the method pertinent to the present work are first summarized; the reader is referred to \citep{schmidt2020} for further details.
Similar to the SPOD algorithm, the LES snapshots are segmented into blocks with a periodic window. Once the Fourier transform is performed, the auto-bispectral matrix is given by
\begin{equation}
\boldsymbol{B_a} = \frac{1}{n_{blk}} \hat{\boldsymbol{Q}}^H_{k \circ l} \boldsymbol{W}  \hat{\boldsymbol{Q}}_{k + l},
\label{eq:bmd1}
\end{equation}
where $\hat{\boldsymbol{Q}}^H_{k \circ l} = \hat{\boldsymbol{Q}}^*_k \circ \hat{\boldsymbol{Q}}^*_l$, $\hat{\boldsymbol{Q}}_{k + l} = \hat{\boldsymbol{Q}}_k + \hat{\boldsymbol{Q}}_l$, and $\boldsymbol{W}$ is the diagonal weight matrix. Next, the optimal expansions $\boldsymbol{\text{a}}_1$ are obtained by maximizing the absolute value of the Rayleigh quotient of $\boldsymbol{B_a}$, that is
\begin{equation}
\boldsymbol{\text{a}}_1 = \text{arg}  \max_{\|a\| = 1} \left| \frac{\boldsymbol{\text{a}} \boldsymbol{B_a}\boldsymbol{\text{a}}} {\boldsymbol{\text{a}}^* \boldsymbol{\text{a}}}  \right| .
\label{eq:bmd2}
\end{equation}
Then the complex mode bispectrum is obtained by
\begin{equation}
\lambda_1(\omega_k, \omega_l) =  \left| \frac{\boldsymbol{\text{a}}_1 \boldsymbol{B_a}\boldsymbol{\text{a}}_1} {\boldsymbol{\text{a}}_1^* \boldsymbol{\text{a}}_1}  \right| .
\label{eq:bmd3}
\end{equation}
The mode bispectrum displays the dominant triadic interaction $(\omega_k, \omega_l, \omega_{k+l})$ (effect) due to the cause $(\omega_k \circ \omega_l)$.

In the present study, 3D $u$-velocity snapshots are utilized, using the same spectral parameters as SPOD analysis with $L2$ weight defined by
\begin{equation}
E = \int_V   u' u'  \,dV.
\label{eq:E_L2}
\end{equation}
BMD is performed for the half-domain, as well as for the separated regions of the USL, LSL, and SPSL, as indicated in figure~\ref{fig:u_mean}(d).

\begin{figure}
     \centering
         \includegraphics[width=0.9\textwidth]{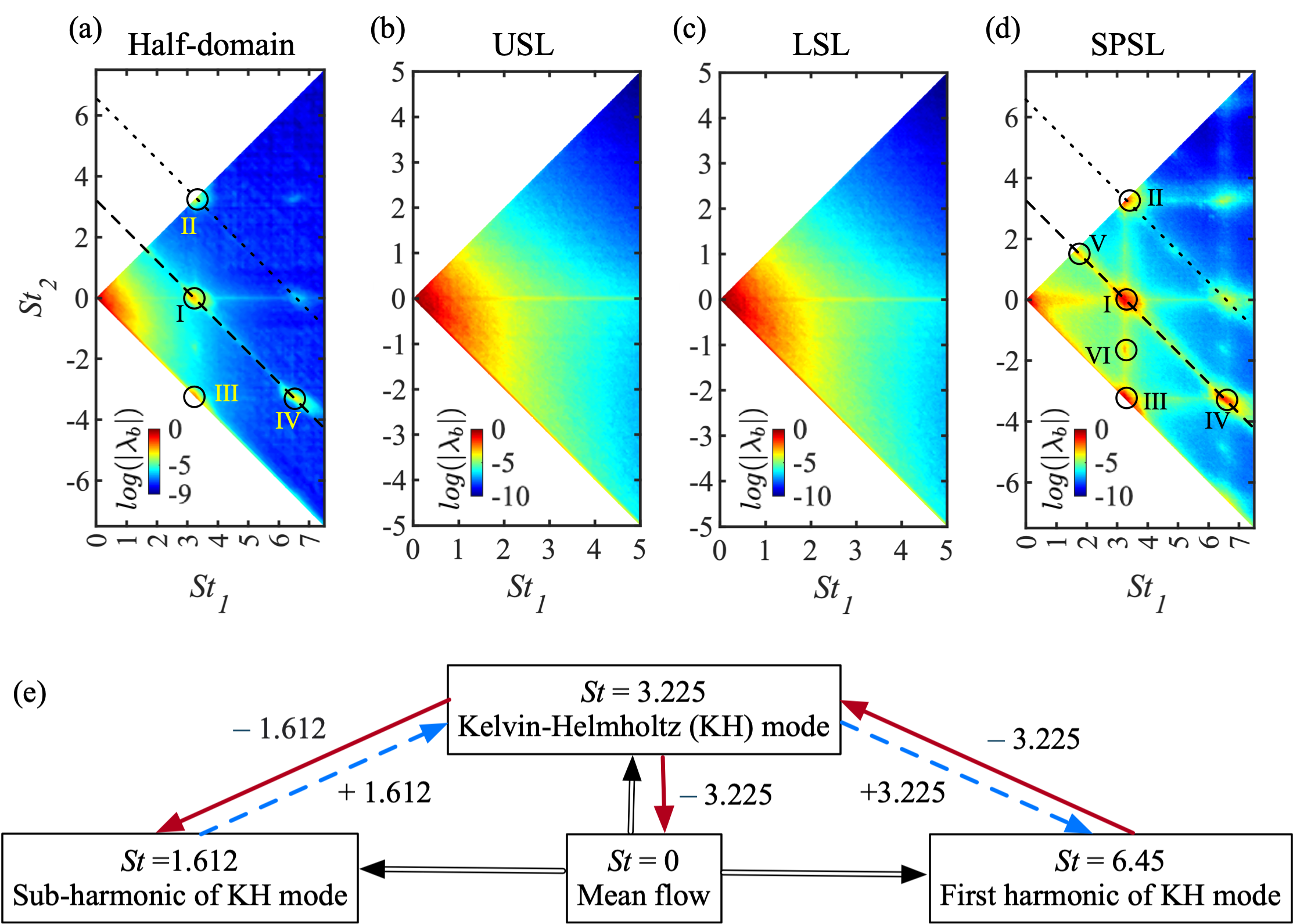}
    \caption{Normalized mode bispectrum $\lambda_b = \lambda_1/| \lambda_1|_{\text{max}}$ for the different cases: (a) half-domain, (b) upper shear layer (USL), (c) lower shear layer (LSL), and (d) splitter plate shear layer (SPSL). The constant frequency $St = 3.225$ and $6.45$ are marked by $--$ and $\dots$ lines, respectively. The circles I-VI indicate the dominant triad interactions. (e) Energy cascade in the SPSL regions for the triads highlighted in the circles. Marron solid and blue dashed arrows show the difference and sum interactions. The double-lined arrow presents the energy cascade from the mean flow.}
    \label{fig:bmd_spectra}
\end{figure}

The normalized mode bispectrum ($\lambda_b = \lambda_1/|\lambda_1|_{\text{max}}$) is shown in figure~\ref{fig:bmd_spectra}. 
The sum and difference interaction regions are given by $St_2>0$ and $St_2<0$, respectively. The diagonal lines present a constant frequency, for example, $St=3.225$ and $St=6.45$ are marked by the dashed and dotted lines, respectively, in figure~\ref{fig:bmd_spectra}(a).  
Similar to the SPOD, the half-domain case combines the effect of all three shear layer regions (figure~\ref{fig:bmd_spectra}(a)).
In the half-domain case, the jet flow exhibits overall broadband behavior at low frequencies $St < 1.5$, indicating that energy transfer predominantly occurs across a wide range of low frequencies.
The dominant tonal frequency at $St = 3.225$ appears as a linear response to the mean flow, marked as region~I in figure~\ref{fig:bmd_spectra}(a).
This frequency exhibits self-interaction through both sum ($3.225, 3.225, 6.45$) and difference ($3.225, -3.225, 0$) mechanisms, resulting in the generation of harmonics (indicated by~II) and mean flow distortion (indicated by~III), respectively.
The bispectrum additionally reveals a strong nonlinear interaction between the first harmonic and the dominant frequency in a difference triad ($6.45, -3.225, 3.225$), denoted by~IV.

Figures~\ref{fig:bmd_spectra}(b) and~\ref{fig:bmd_spectra}(c) show the normalized mode bispectrum for the USL and LSL regions, respectively. In both cases, broadband behavior is observed, with strong nonlinear interactions concentrated at low frequencies ($St < 1.5$). The presence of a prominent band along $St_2 = 0$ indicates energy exchange involving the mean flow, contributing to the energy cascade.
In contrast, the SPSL bispectrum (figure~\ref{fig:bmd_spectra}(d)) exhibits a primarily tonal behavior, characterized by energy transfer through sum and difference interactions of the dominant frequency $St = 3.225$. In addition to the previously noted hotspots (I–IV) from figure~\ref{fig:bmd_spectra}(a), two additional nonlinear triadic interactions are observed: a sum interaction (V), corresponding to ($1.612, 1.612, 3.225$), and a difference interaction (VI), corresponding to ($3.225, -1.612, 1.612$).

The overall energy cascade in the SPSL region, involving the fundamental KH instability and its sub-harmonic and first harmonic interactions, is illustrated in figure~\ref{fig:bmd_spectra}(e).
These triadic interactions emphasize the critical role of quadratic nonlinearities in redistributing energy across frequencies in the jet flow. These findings are consistent with previous investigations of the 2D isolated shear layer, which similarly revealed a cascade of nonlinear triadic interactions originating from the fundamental KH instability \citep{thakor2024}.

\section{Triglobal resolvent analysis} \label{sec:resl}
Although SPOD is highly effective for extracting coherent structures, it requires long, time-resolved datasets, making it computationally demanding for fully 3D flows.
Hence, it becomes essential to model coherent structures.
The resolvent model provides an input–output transfer function that identifies the most amplified flow responses to given perturbations, offering insight into coherent structure formation and amplification mechanisms \citep{reddy1993, trefethen1993, jovanovic2005}.
The time-averaged mean of a turbulent flow can be used as a base state about which the Navier-Stokes (N-S) equations are linearized.
In this framework, the N-S equations are interpreted as a forced linear system about the turbulent mean flow, with the nonlinear terms acting as an intrinsic forcing \citep{mckeon2010}.
Previous studies have successfully used resolvent analysis to identify amplification mechanisms in jet flows \citep{garnaud2013, jeun2016, schmidt2018, thakor2025}.

\subsection{Formulation and configuration}
The governing equations for perturbations, $\boldsymbol{q}'$, are obtained from the 3D compressible N-S equations and written in an input-output framework
\begin{equation}
    \frac{d {\boldsymbol q'}}{dt} = \boldsymbol {L}(\boldsymbol{\overline{q}}){\boldsymbol q'} + \boldsymbol{B f'},
    \label{eq:IO1}
\end{equation}
\begin{equation}
\boldsymbol{y_o} = \boldsymbol{C q'},
    \label{eq:IO2}
\end{equation}
where $\boldsymbol {L}(\boldsymbol{\overline{q}})$ denotes the linearized compressible N-S operator about the 3D mean state $\boldsymbol{\overline{q}}(x,y,z)$.
All three spatial directions are treated as inhomogeneous, consistent with the triglobal framework.
Here, $\boldsymbol f'$ is considered the nonlinear perturbation term as well as an external forcing \citep{mckeon2010}).

${\boldsymbol q}'$ and ${\boldsymbol f}'$ are expressed in terms of their Fourier representation as
\begin{equation}
  [ \boldsymbol{q}' (\boldsymbol{x},t) ; \boldsymbol{f}' (\boldsymbol{x},t)  ]  =
 \int_{-\infty} ^{\infty}  [ \boldsymbol {\hat{q}(x)};   \boldsymbol {\hat{f}(x)} ] e^{- i \omega t} d\omega.
    \label{eq:fourier}
\end{equation}
where $\boldsymbol{x}$ is a spatial vector, i.e., $\boldsymbol{x} = [x,y,z]$. The above system of equations yields in the frequency domain
\begin{equation}
   \boldsymbol {\hat{y}_o}  = \boldsymbol{C}[ -i \omega \boldsymbol{I} - \boldsymbol{L}(\boldsymbol {\overline{q}})]^{-1} \boldsymbol{B} \hat{\boldsymbol f} = {\boldsymbol{H}}(\boldsymbol {\overline{q}}; \omega) \hat{\boldsymbol f},
\end{equation}
where ${\boldsymbol{H}}(\boldsymbol{\overline{q}}; \omega) = \boldsymbol{C}[ -i \omega \boldsymbol{I} - \boldsymbol{L}(\boldsymbol {\overline{q}})]^{-1} \boldsymbol{B} $ is known as the {\it resolvent operator}.
The operator $\boldsymbol {H}(\boldsymbol{\overline{q}}; \omega)$ is a transfer function from the sustained input variable $\hat{\boldsymbol f}$ to the harmonic output variable $\boldsymbol {\hat{y}_o}$ associated with a frequency $\omega$ for the given 3D mean state $\boldsymbol {\overline{q}}$.

The optimization problem to seek maximum gain between inputs and outputs \citep{schmid2002} is formulated as follows:
\begin{equation}
    G^2_{\text{max}} = \max_{\boldsymbol{\hat {f}}} \frac{\langle \boldsymbol{{\hat {y}}_o}, \boldsymbol{{\hat {y}}_o}\rangle_E }{\langle {\boldsymbol{\hat {f}}}, {\boldsymbol{\hat{f}}} \rangle_E } = \frac{\boldsymbol{{\hat {y}}_o}^* \boldsymbol{W_y} \boldsymbol{{\hat {y}}_o}}{\boldsymbol{\hat{f}}^* \boldsymbol{W_f} {\boldsymbol{\hat{f}} } } = \| \boldsymbol{W_y}^{1/2} \boldsymbol{H}(\boldsymbol {\overline{q}}; \omega) \boldsymbol{W_f}^{-1/2} \|^2_2= \sigma_1^2.
    \label{eq:max_g}
\end{equation}
Hence, the weighted resolvent operator $ \boldsymbol{R}(\boldsymbol {\overline{q}}; \omega) = \boldsymbol{W_y}^{1/2} \boldsymbol{H}(\boldsymbol {\overline{q}}; \omega) \boldsymbol{W_f}^{-1/2}$ can be cast in the framework of singular value decomposition (SVD) to determine the harmonic forcing $\hat{\boldsymbol{f}}$ and the corresponding response $\boldsymbol{\hat{y}_o}$. The decomposition of the resolvent operator is given by
\begin{equation}
\boldsymbol{R}(\boldsymbol {\overline{q}}; \omega) = \boldsymbol{\mathcal{U}_y \Sigma \mathcal{V}_f}^*,
\end{equation}
where, $\boldsymbol{\mathcal{U}_y}  = [\hat {\boldsymbol y}_1,\hat {\boldsymbol y}_2,\dots,\hat {\boldsymbol y}_k]$ is a set of the orthogonal response modes $\hat {\boldsymbol y}_j$, and $\boldsymbol{\mathcal{V}_f} = [\hat {\boldsymbol f}_1,\hat {\boldsymbol f}_2,\dots, \hat {\boldsymbol f}_k]$ contains a set of the orthogonal forcing modes $\hat {\boldsymbol f}_j$ \citep{trefethen1993, schmid2002}. The diagonal matrix $\boldsymbol{\Sigma} = \text{diag}(\sigma_1,\sigma_2,\dots,\sigma_k)$ yields amplitude gain, with $\sigma_k^2$ representing the energy amplification ratio of the response and forcing modes, depending on a norm specified.

In the present work, the resolvent gain is quantified using the compressible energy defined in Eq.~(\ref{eq:E_chu}) for both inputs and outputs, i.e., $\boldsymbol{W_y} = \boldsymbol{W_f} = \boldsymbol{W}$.
The input matrix $\boldsymbol{B}$ and the output matrix $\boldsymbol{C}$ are applied to perform the analysis in specific shear layer regions. The USL and LSL are considered within domains $(x/w_{sp},y/w_{sp},z/w_{sp}) \in [(0, 0.16, -0.7) \times (1.38, 0.6, 0)]$ and $[(1.09, -0.25, -0.7) \times (2.6, 0.25, 0)]$, respectively, to isolate the amplification mechanisms in these regions.
Outside these domains, $\boldsymbol{B}$ and $\boldsymbol{C}$ are set to zero for the corresponding cases.
We focus on the USL and LSL regions for the triglobal resolvent analysis because the instability mechanisms there are highly 3D, as shown in the SPOD analysis (\S\ref{sec:spod}).
The SPSL region, on the other hand, is primarily governed by 2D mechanisms, which have already been studied extensively in our companion works \citep{thakor2024, thakor2025}).

An in-house code was first developed for the 3D linearized compressible Navier–Stokes operator, ($\boldsymbol{L}(\boldsymbol{\overline{q}})$) in the \textit{CharLES} \citep{bres2017} framework.  
The in-house code is validated by eigen-decomposition analysis for a canonical 3D cubic lid-driven cavity flow (for completeness, a summary is provided in Appendix~\ref{sec:app_3d_lin_op}).
The linear operator $\boldsymbol{L}(\boldsymbol{\overline{q}})$ for the 3D jet flow is then obtained by interpolating the LES time-averaged flow onto a mesh prepared suited for resolvent analysis.
The operator is discretized on a structured 3D grid with approximately $0.32$ million points in each region.
Zero gradient Neumann boundary conditions are prescribed for the fluctuating pressure, and Dirichlet boundary conditions are applied to the velocity components (i.e. $u'_i = 0$) at all boundaries, except at the nozzle center plane ($z=0$), where a symmetry condition is imposed for all the variables~\citep{sun2017, ribeiro2023, thakor2024}.
Sponge zones are applied at all outflow boundaries to damp perturbations and reduce the influence of boundary effects on the inner-domain results. The Reynolds number is set to $1{,}000$ to construct the operator.

For each case, the resulting triglobal resolvent operator has a size of approximately $1.6 \times 10^6$ in a square matrix form.
Since performing a direct SVD on such a large matrix operator is computationally expensive, the randomized algorithm described by~\citep{ribeiro2020} is employed.
Based on convergence analysis of the test vectors and power iterations (see Appendix~\ref{sec:app_rsvd}), $10$ test vectors are sufficient to reduce computational cost without compromising accuracy.
The computational cost of obtaining the randomized SVD at a single frequency is approximately between $3$ to $4$ TB of memory and $24$ to $35$ CPU hours, depending on input frequency.
Hence, the triglobal resolvent analysis is performed over the frequency range $0.375 \leq St \leq 3.5$, with a finer resolution of $\Delta St = 0.0625$ up to $St = 1$ and a coarser resolution of $\Delta St = 0.25$ up to $St = 3.5$.
At lower frequencies ($St < 0.375$), domain truncation effects arise because the modal structures become highly elongated.

\subsection{Resolvent spectra and modes}

\begin{figure}
     \centering
         \includegraphics[width=0.85\textwidth]{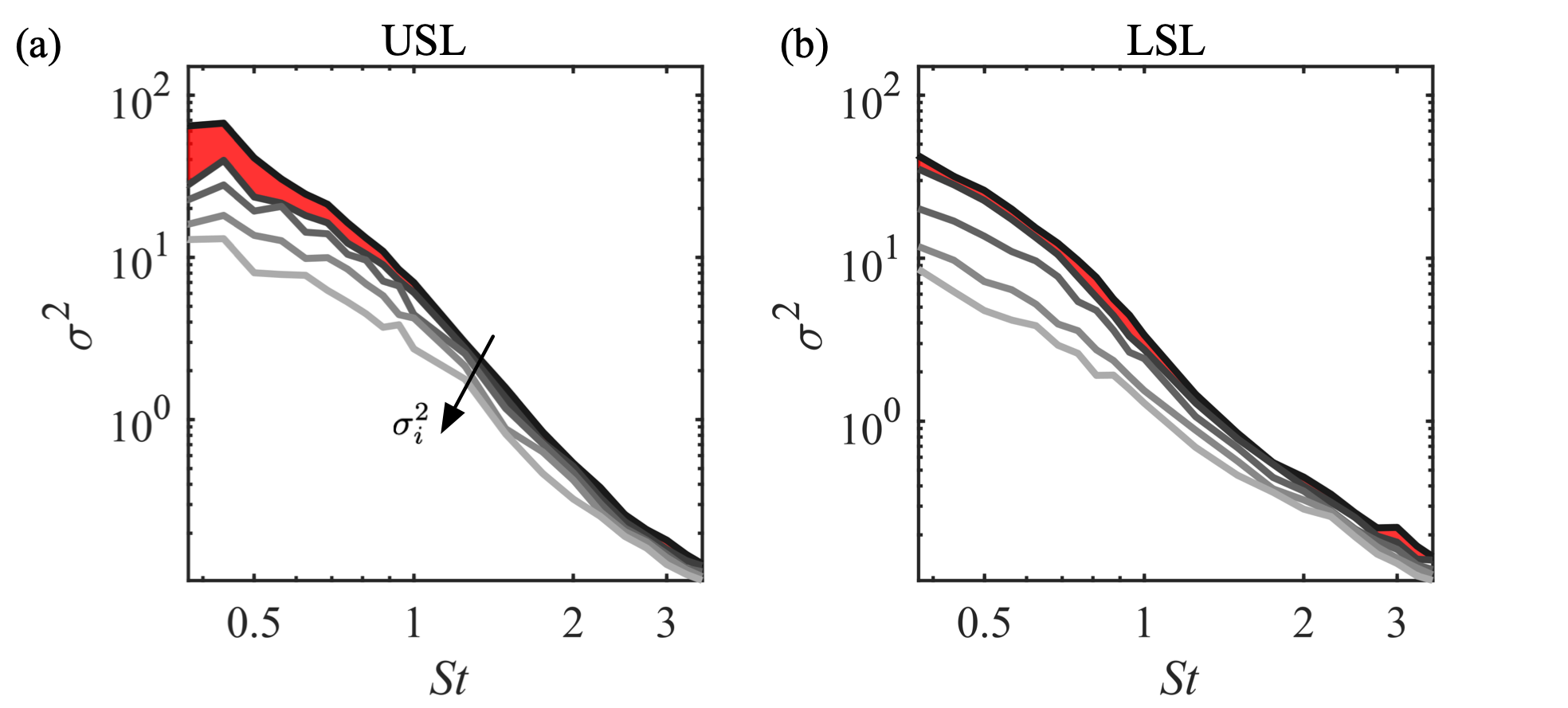}
    \caption{The leading five resolvent amplification energy ($\sigma^2$) based on the compressible energy weight for: (a) upper shear layer (USL), and (b) lower shear layer (LSL). At each frequency, decreasing energy gains are shown in lighter shades, i.e., $\sigma^2_1 \geq \sigma^2_2 \dots \geq \sigma^2_5$. The red-shaded area highlights the separation between the optimal ($\sigma^2_1$) and the first sub-optimal ($\sigma^2_2$) singular values.}
    \label{fig:resl_spectra}
\end{figure}

Figures~\ref{fig:resl_spectra}(a) and~\ref{fig:resl_spectra}(b) display the resolvent amplification energy spectra, computed based on the compressible energy norm, for the USL and LSL regions, respectively.
In both cases, higher resolvent gains are observed at low frequencies, with a sharp decay as the frequency increases.
For the USL (figure~\ref{fig:resl_spectra}(a)), the optimal gain ($\sigma_1$) is clearly separated from the first sub-optimal gain ($\sigma_2$) at low frequencies ($St \leq 1$), as highlighted by the red-shaded region.
In contrast, no significant separation between $\sigma_1$ and $\sigma_2$ is observed in the LSL region (figure~\ref{fig:resl_spectra}(b)). The resolvent spectra exhibit maximum values at $St = 0.4375$ and $St = 0.375$ for the USL and LSL, respectively.
Overall, the trends in the resolvent amplification energy are similar to the corresponding SPOD spectra, shown earlier in figure~\ref{fig:spod_spectra}.
This indicates that the linear amplification mechanism remains active in the complex 3D jet flow.

\begin{figure}
     \centering  \includegraphics[width=0.95\textwidth]{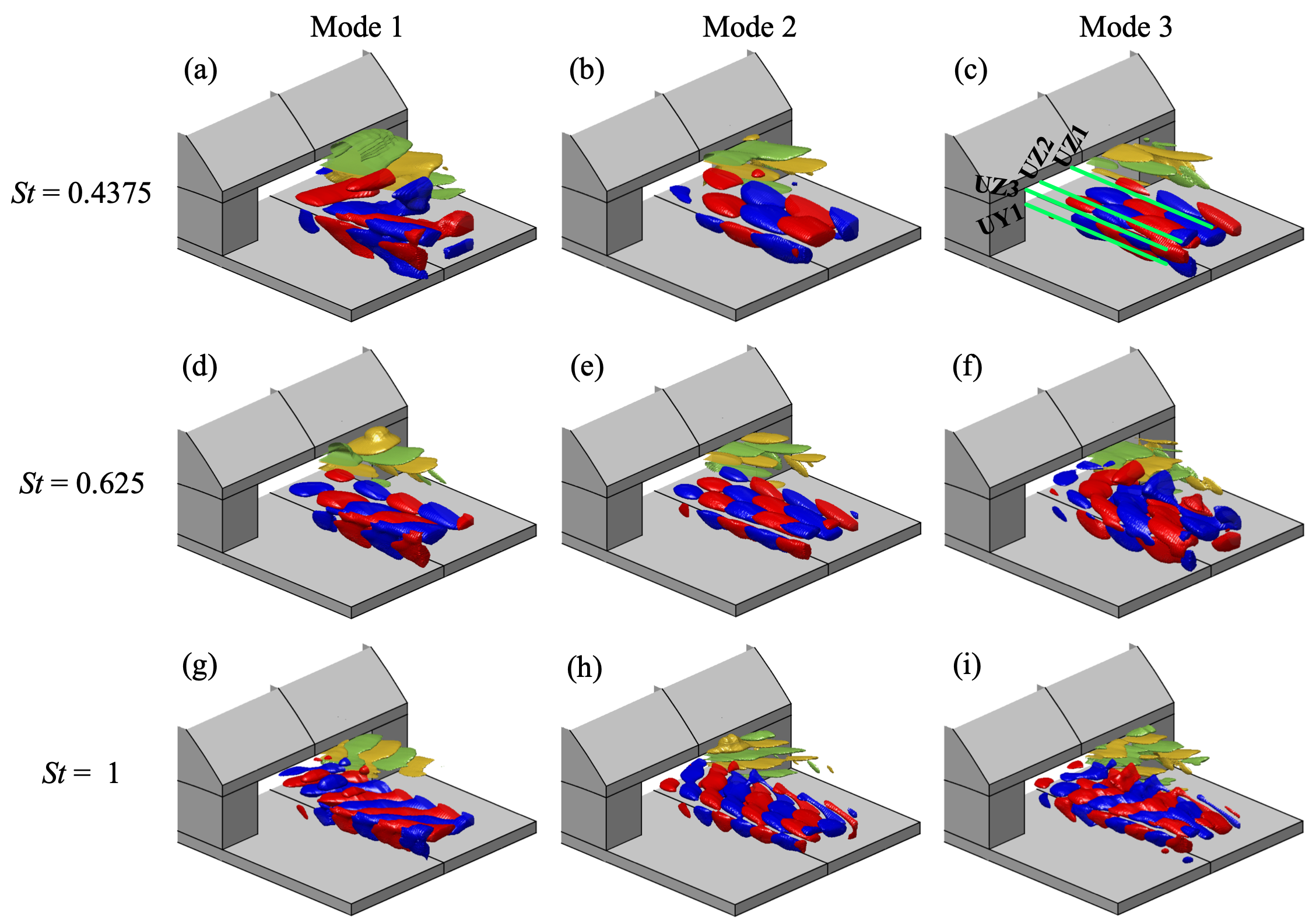}
    \caption{The leading three resolvent mode pair for the upper shear layer (USL) region. (a-c) $St = 0.4375$, (d-f) $0.625$, (g-i) $0.4375$. Iso-surface of the real component of $\boldsymbol{\hat{u}}/\|\boldsymbol{\hat{u}}\|_{\text{max}}$ of  Response: \textcolor{blue}{\rule{1.5ex}{1.5ex}}\textcolor{red}{\rule{1.5ex}{1.5ex}} (blue, red) = [-0.2, 0.2], and forcing:  \textcolor{green}{\rule{1.5ex}{1.5ex}}\textcolor{yellow}{\rule{1.5ex}{1.5ex}} (green, yellow) = [-0.2, 0.2]. The green lines marked in (c) are used for analysis in the following analysis.}
    \label{fig:resl_usl_modes}
\end{figure}

Figure~\ref{fig:resl_usl_modes} displays the optimal and two leading sub-optimal resolvent forcing (green-yellow iso-surfaces) and response (blue-red iso-surfaces) modes for the USL region.
Overall, the forcing modes are localized near the nozzle exit across all low frequencies ($St \leq 1$), while the corresponding response modes emerge further downstream within the shear layer.
This spatial separation reflects the convective nature of the KH instability that originates in the initial shear layer of the developing jet \citep{crow1971, krothapalli1986, schmidt2018}).

At the frequency of maximum gain ($St = 0.4375$), the optimal and sub-optimal forcing–response mode pairs exhibit distinct spatial structures, as shown in figure~\ref{fig:resl_usl_modes}(a–c).
The optimal response mode is characterized by oblique structures occupying the upper and side shear layers, while the sub-optimal response modes display highly 3D, spanwise-alternating wavepackets.
The optimal high-amplitude response modal structures appear in the corner region, which is further explained in the following \S\ref{sec:corner}.
At $St = 0.625$, the optimal response mode features a combination of oblique and spanwise-alternating structures (figure~\ref{fig:resl_usl_modes}(d)), whereas the sub-optimal response modes primarily retain spanwise-alternating wavepackets (figure~\ref{fig:resl_usl_modes}(e–f)). 
At the moderately higher frequency of $St = 1$, both optimal and sub-optimal response modes exhibit oblique wavepackets with noticeably shorter wavelengths (figure~\ref{fig:resl_usl_modes}(g–i)).
As the frequency increases from $St = 0.4375$ to $1$, the optimal response modal structures shift from the corner region toward the center plane of the upper shear layer, which is further discussed in \S\ref{sec:corner} and \S\ref{sec:wavemaker}.

\begin{figure}
     \centering
         \includegraphics[width=0.95\textwidth]{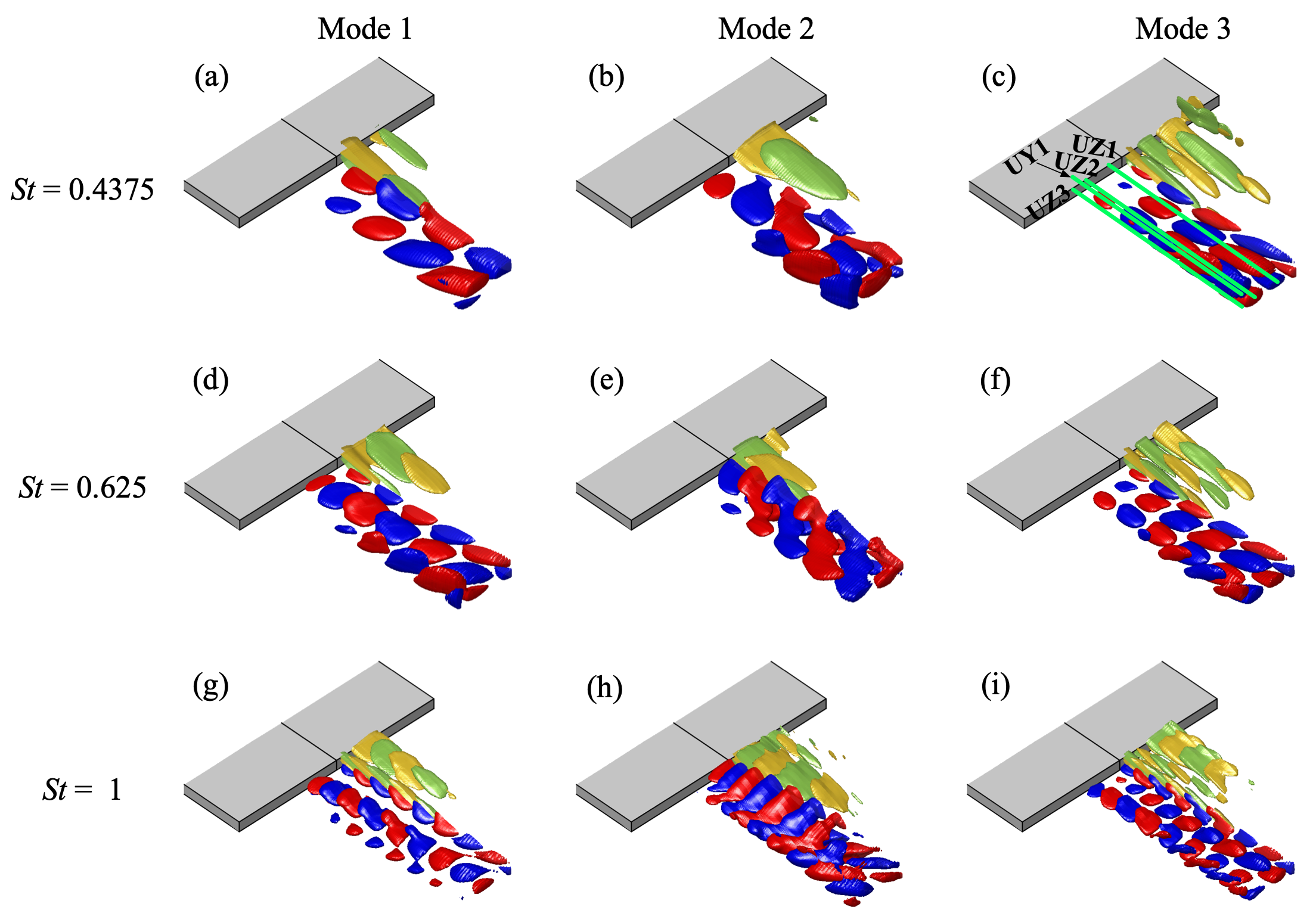}
    \caption{The leading three resolvent mode pair for the lower shear layer (LSL) region. (a-c) $St = 0.4375$, (d-f) $0.625$, (g-i) $1$. Iso-surface of the real component of $\boldsymbol{\hat{u}}/\|\boldsymbol{\hat{u}}\|_{\text{max}}$ of  Response: \textcolor{blue}{\rule{1.5ex}{1.5ex}}\textcolor{red}{\rule{1.5ex}{1.5ex}} (blue, red) = [-0.2, 0.2], and forcing:  \textcolor{green}{\rule{1.5ex}{1.5ex}}\textcolor{yellow}{\rule{1.5ex}{1.5ex}} (green, yellow) = [-0.2, 0.2]. The green lines marked in (c) are used for analysis in the following analysis.}
    \label{fig:resl_lsl_modes}
\end{figure}

Next, the optimal and sub-optimal resolvent modes in the LSL region are presented in figure~\ref{fig:resl_lsl_modes}, focusing on the region downstream of the aft deck plate.
Similar to the USL case, the forcing modes appear upstream of the response modes, consistent with the convective nature of KH instabilities in the developing region of the LSL.
Spanwise-alternating structures are observed in both optimal and sub-optimal response modes across the examined frequencies.
At $St = 0.4375$, the optimal response mode is concentrated near the center plane (figure~\ref{fig:resl_lsl_modes}(a)), while the sub-optimal response modes exhibit structures extending toward the side edges of the shear layer (figure~\ref{fig:resl_lsl_modes}(b–c)).
Mode switching is observed at $St = 0.625$, the first sub-optimal response mode exhibits center-plane dominance (figure~\ref{fig:resl_lsl_modes}(e)).
At $St = 1$, the optimal response mode displays 3D structures near the center plane, while the first sub-optimal response mode exhibits a predominantly two-dimensional response (figure~\ref{fig:resl_lsl_modes}(h)).
Since a clear separation between the optimal and sub-optimal gain is not observed (figure~\ref{fig:resl_spectra}(b)), multiple amplification mechanisms remain comparatively active in this region.
Hence, the sub-optimal modes are equally necessary to capture the underlying flow physics accurately.

The resolvent high-amplitude modal spatial structures tend to localize either near the nozzle corner or close to the center plane.
To investigate the physical amplification mechanism in the center and nozzle corner, the Fourier transform is applied to the streamwise line data of optimal streamwise velocity response mode along the four selected locations for both the USL and LSL regions.
For the USL, the selected $(y/w_{sp},z/w_{sp})$ locations, as shown in figure~\ref{fig:resl_usl_modes}(c), are UZ1 = $(0.357, -0.1) $, UZ2 = $(0.357, -0.4) $, UZ3 = $(0.357, -0.5) $ and UY1 = $(0.257, -0.5)$, where $y/w_{sp}=0.357$ and $z/w_{sp}=-0.5$ are the nozzle lip line along the major and minor axis, respectively. Similarly, in the LSL region, points are chosen along the aft-deck upper surface line ($y/w_{sp}=0$) at UZ1 = $(0, -0.1) $, UZ2 = $(0, -0.4) $, UZ3 = $(0, -0.5) $ and UY1 = $(0.1, -0.5)$, as shown in figure~\ref{fig:resl_lsl_modes}(c).

\begin{figure}
     \centering
         \includegraphics[width=1\textwidth]{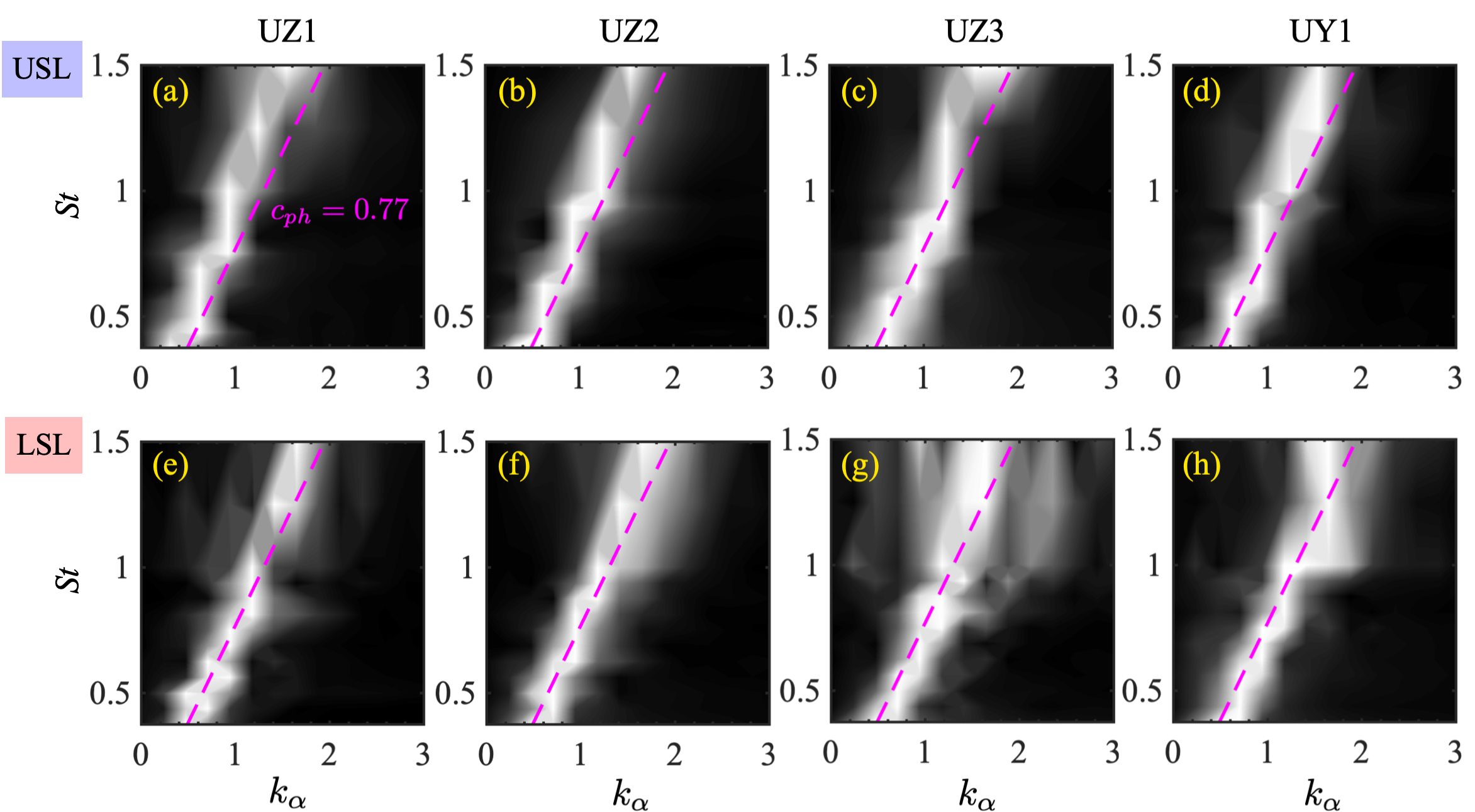}
    \caption{The normalized spatio-temporal power spectral density (PSD) of the optimal $u$-velocity response mode at four representative locations. (a-d) upper shear layer (USL) region, (e-h) lower shear layer (LSL) region. $P_u/\|P_u\|_{\text{max}}$:  \fcolorbox{black}{black}{}\fcolorbox{black}{white}{} (black, white) $\in$ [0, 1]. The magenta dashed (- -) line presents a constant phase speed $c_{ph} =0.77$.}
    \label{fig:resl_phase_speed}
\end{figure}

The spatio-temporal power spectral density (PSD) is computed as $P_u = |\boldsymbol{\hat{u}}_{\omega k_\alpha}|^2$ by applying a Fourier transform in the streamwise direction, where $k_\alpha$ denotes the streamwise wavenumber.
The normalized PSD is evaluated at the above-mentioned spatial locations in both the USL and LSL and is shown in figure~\ref{fig:resl_phase_speed}.
The results exhibit qualitatively similar trends across all locations in the USL (figure~\ref{fig:resl_phase_speed}(a--d)) and LSL (figure~\ref{fig:resl_phase_speed}(e--h)).
The PSD of the streamwise velocity component aligns along a constant phase speed line, $c_{ph} = 0.77 $, which is typically associated with KH-type shear layer instability waves \citep{crow1971, gudmundsson2011, jordan2013, schmidt2018}.
For $St > 1$, the spectra in both the USL and LSL regions exhibit broadband features, as a result of the dominant low-frequency dynamics in these regions, while the amplification mechanism remains predominantly governed by the KH-type wavepackets originating within the initial shear layer.

\subsection{Corner effects in rectangular jet} \label{sec:corner}

One of the key features of rectangular or non-axisymmetric jets is the presence of complex 3D phenomena associated with corner regions \citep{krothapalli1981, ho1987, hussain1989}.
The formation of streamwise vortices near the corners can either entrain ambient fluid into the potential core or eject core fluid outward. This corner-induced mechanism plays a crucial role in either promoting or resisting axis-switching in rectangular jets \citep{zaman1996}.

\begin{figure}
     \centering
         \includegraphics[width=1\textwidth]{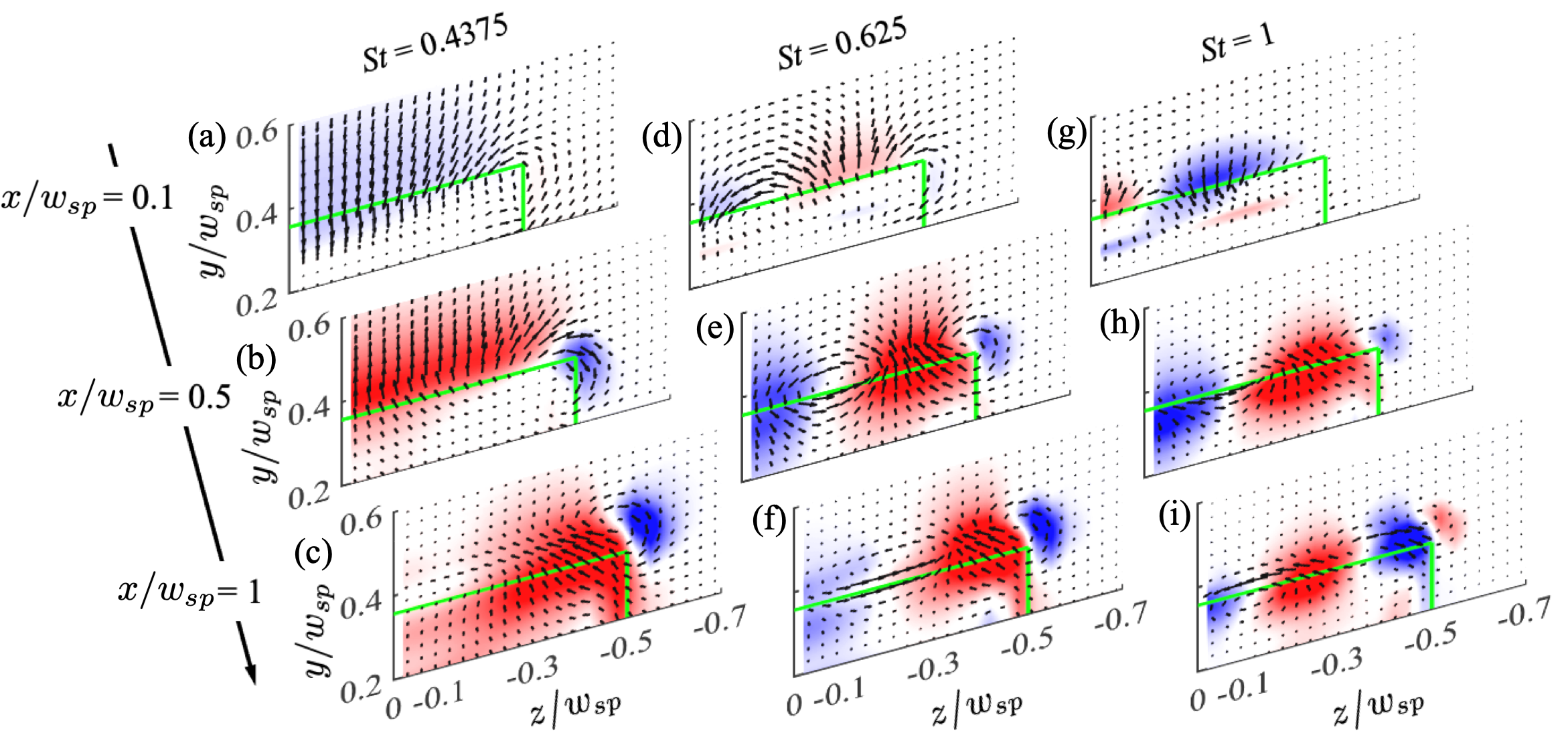}
    \caption{Optimal response mode at various streamwise $y-z$ slices for representative frequencies for the USL region. (a-c) $St = 0.4375$, (d-f) $0.625$, and (g-i) $1$. The real component of $\boldsymbol{\hat{v}}/\|\boldsymbol{\hat{v}}\|_{\text{max}}$ of  Response: \textcolor{blue}{\rule{1.5ex}{1.5ex}}\textcolor{red}{\rule{1.5ex}{1.5ex}} (blue, red) $\in$ [-0.2, 0.2]. The green solid line represents the nozzle lip line.}
    \label{fig:resl_usl_corner}
\end{figure}

To further examine these corner-induced effects, figure~\ref{fig:resl_usl_corner} presents 2D streamwise $x-$normal slices illustrating the $v$-velocity contours overlapped with planar-restricted velocity vectors of the optimal response mode.
These vectors indicate the presence of a strong vortex near the corner at $St = 0.4375$ close to the nozzle exit (nozzle exit plane is at $x/w_{sp} =0$.) (see figure~\ref{fig:resl_usl_corner}(a)).
This vortex entrains the ambient fluid along the major axis and ejects the jet flow to the ambient along the minor axis.
Moving downstream to $x/w_{sp} = 0.5$, the corner vortex reverses its rotation direction and continues to convect with the same sense of rotation, thereby altering the direction of entrainment and ejection.
As the forcing frequency increases, the strength of the corner vortex decreases, as shown in figure~\ref{fig:resl_usl_corner}(d-i). At higher frequencies, such as $St = 1$, the entrainment and ejection predominantly occur along the major axis, indicating a reduced influence of corner-induced effects.

\begin{figure}
     \centering
         \includegraphics[width=1\textwidth]{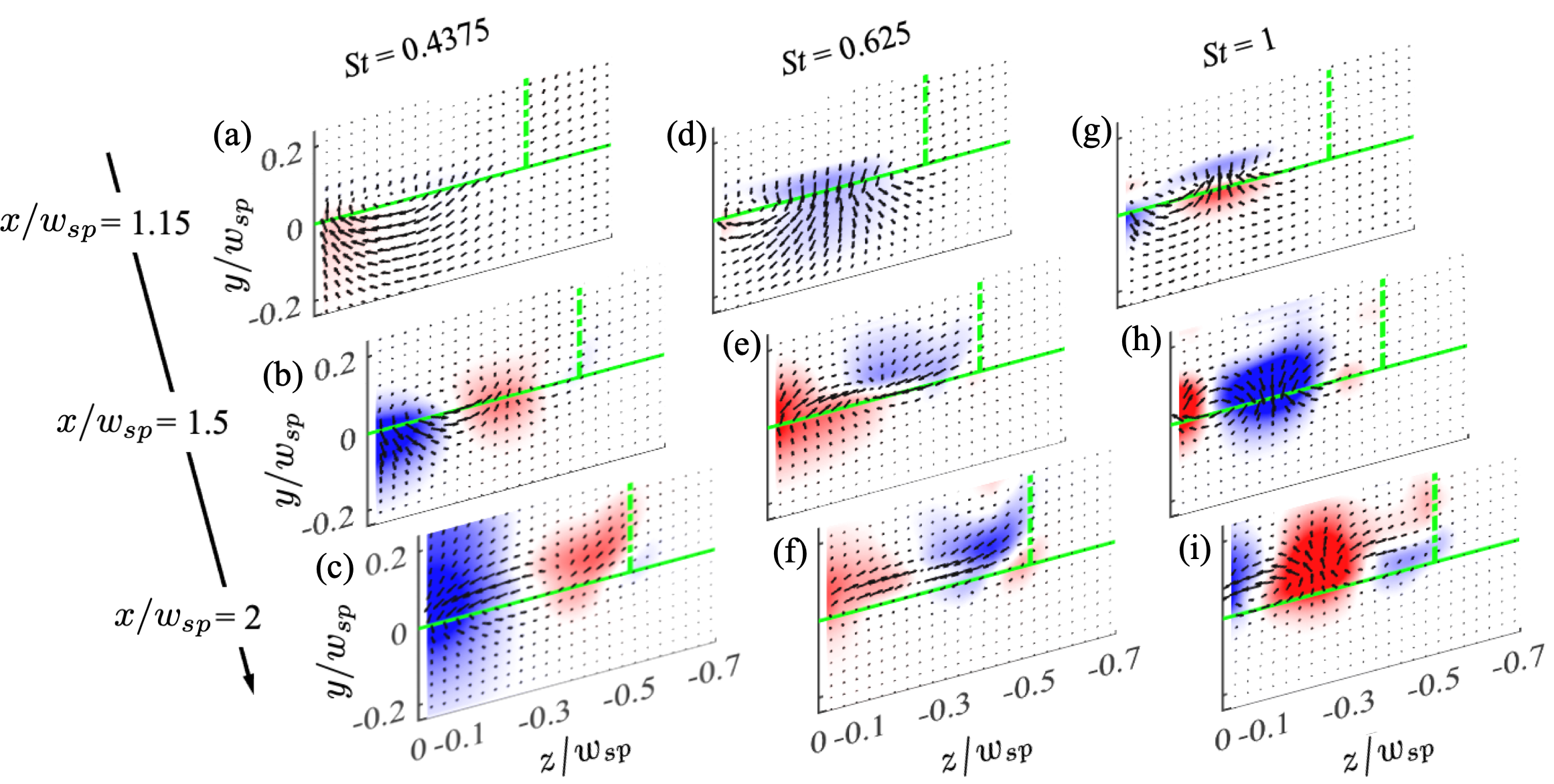}
    \caption{The optimal response mode at various streamwise $y-z$ slices for representative frequencies for the LSL region. (a-c) $St = 0.4375$, (d-f) $0.625$, and (g-i) $1$. The real component of $\boldsymbol{\hat{v}}/\|\boldsymbol{\hat{v}}\|_{\text{max}}$ of  Response: \textcolor{blue}{\rule{1.5ex}{1.5ex}}\textcolor{red}{\rule{1.5ex}{1.5ex}} (blue, red) $\in$ [-0.2, 0.2]. The green vertical dashed and horizontal solid lines represent the nozzle lip and aft-deck upper surface edge, respectively.}
    \label{fig:resl_lsl_corner}
\end{figure}

Figure~\ref{fig:resl_lsl_corner} displays 2D streamwise $x-$normal slices of velocity vectors and $v$-velocity contours for the optimal response mode in the LSL region.
In contrast to the USL case, no distinct corner vortex forms in the LSL, which is likely due to the wider aft-deck representing the integration of the nozzle to the aircraft frame; strong vortical structures are instead observed near the center plane ($z = 0$).
At $St = 0.4375$, a prominent vortex appears near the trailing edge of the aft-deck ($x/w_{sp} = 1.15$), entraining fluid from the side shear layer and lifting the lower shear layer toward the center. As the flow convects downstream, vortex pairs intensify and propagate upward from the aft-deck, as shown in figures~\ref{fig:resl_lsl_corner}(b-c). This upward displacement of the lower shear layer contributes to a reduction in the potential core length \citep{stack2018}. Higher-frequency cases, such as $St = 0.625$ and $1$, show qualitatively similar flow dynamics further downstream, as displayed in figures~\ref{fig:resl_lsl_corner}(d-i).

\subsection{Resolvent wavemaker} \label{sec:wavemaker}

Self-excited flow oscillations, such as vortex shedding near the airfoil trailing edge, aircraft buffet, and tonal behavior in jets, arise in regions where perturbations strongly excite and are highly receptive to feedback mechanisms. A \textit{wavemaker}, originally formulated through spatial overlaps of direct and adjoint global eigenmodes in linear stability analysis, captures such sensitive regions in the flow field \citep{strykowski1990, giannetti2007}.

Recent studies have extended this idea through the construction of a \textit{resolvent wavemaker} based on the resolvent modes \citep{qadri2017, skene2022, ribeiro2023, houtman2023}. The resolvent wavemaker quantifies regions of maximal spatial overlap between forcing and response modes, hence pinpointing locations where the flow inherently supports a self-sustained feedback loop. The resolvent wavemaker provides a powerful tool for designing active flow control strategies.

The resolvent wavemaker gain is defined as
\begin{equation}
\zeta = \sigma^2 \int_V |\boldsymbol{\hat{f}} \circ \boldsymbol{\hat{q}} | dV,
\label{eq:wave_3D}
\end{equation}
where $\boldsymbol{\hat{f}} \circ \boldsymbol{\hat{q}}$ is the resolvent wavemaker mode given by the Hadamard product between forcing $\boldsymbol{\hat{f}}$ and response $\boldsymbol{\hat{q}}$ modes. The 2D resolvent wavemaker gain ($\zeta_z$) is computed over $z$-normal slices $S(x,y)$ to identify the spanwise distribution of regions that support self-sustained oscillatory dynamics.

\begin{figure}
     \centering
         \includegraphics[width=0.9\textwidth]{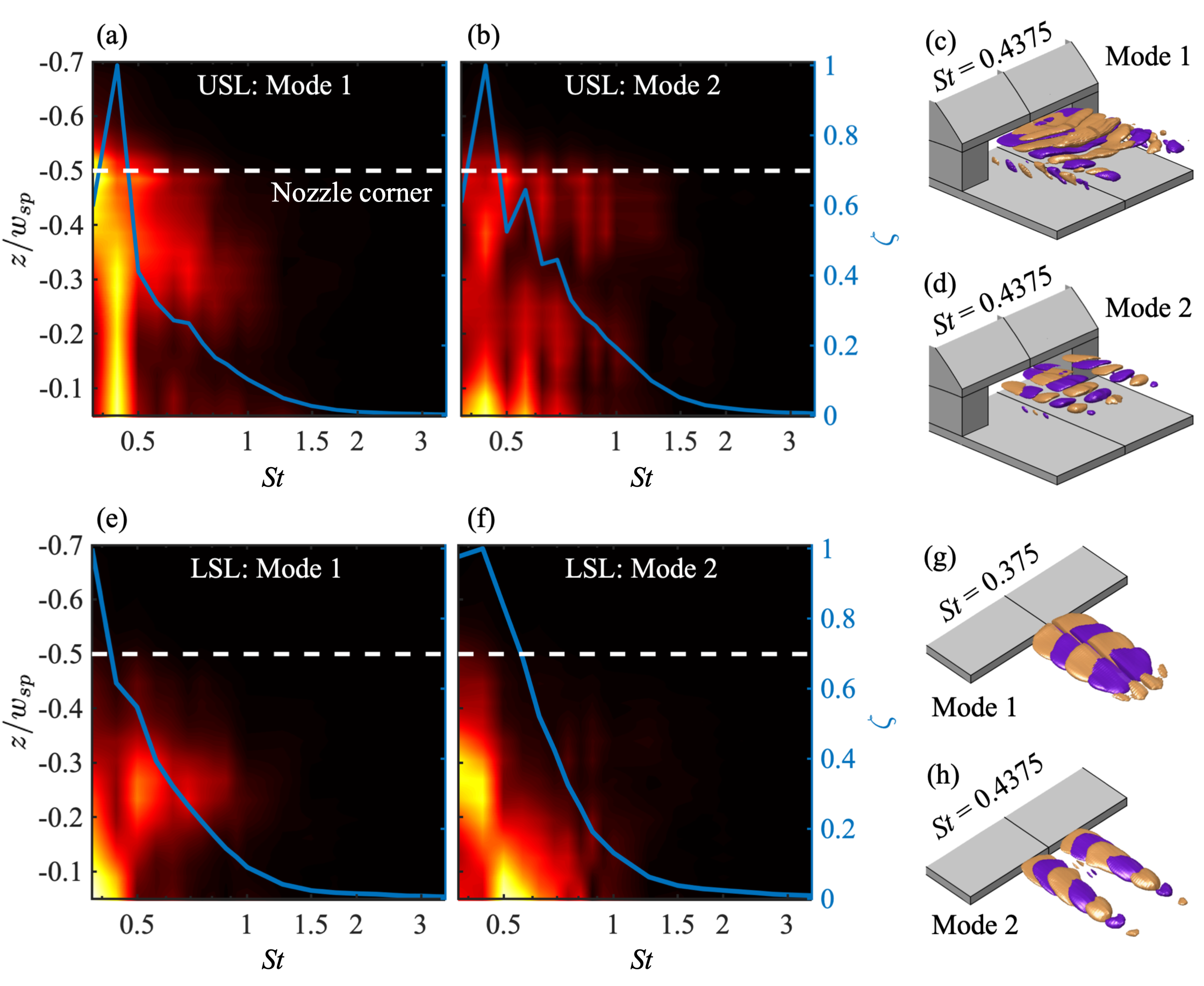}
    \caption{Normalized resolvent wavemaker $\zeta$ (by its maximum value) and contour of $\zeta_z$ computed over each $z$-normal slices for the optimal and suboptimal resolvent modes. (a-d) upper shear layer (USL), (e-h) lower shear layer (LSL) regions. Contour of $\zeta_z$ \textcolor{black}{\rule{1.5ex}{1.5ex}}\textcolor{red}{\rule{1.5ex}{1.5ex}}\textcolor{yellow}{\rule{1.5ex}{1.5ex}}  (black, red, yellow) $\in$ [0, 1]. Resolvent wavemaker mode of the real component of $u$-velocity: \textcolor{violet}{\rule{1.5ex}{1.5ex}}\textcolor{orange}{\rule{1.5ex}{1.5ex}} (purple, orange) $\in$ [-0.2, 0.2]. The white dashed line represents the nozzle corner location at $z/w_{sp}=-0.5$.}
    \label{fig:resl_wavemaker}
\end{figure}

The normalized resolvent wavemaker gains are presented in figure~\ref{fig:resl_wavemaker}(a-b) for the USL and~\ref{fig:resl_wavemaker}(e-f) LSL regions.
We first discuss the wavemaker analysis in the USL region.
The optimal and first sub-optimal wavemaker gains $\zeta$ peak at $St = 0.4375$.
The resolvent spectrum also shows maximum amplification at the same frequency for the optimal and the first sub-optimal modes (see figure~\ref{fig:resl_spectra}(a)).
The optimal wavemaker gain $\zeta_z$, evaluated on each spanwise slice, localizes near the nozzle corner ($z/w_{sp} = -0.5$) for frequencies $St \leq 0.5$.
An exception occurs at $St = 0.4375$, where the gain exhibits spanwise spreading from the nozzle center plane to the corner, as shown by the resolvent wavemaker mode in figure~\ref{fig:resl_wavemaker}(c).
The wavemaker in this region supports the generation of unsteady corner vortices, which play a critical role in promoting axis-switching phenomena in rectangular jets, as discussed in \S\ref{sec:corner}.
Conversely, the first sub-optimal wavemaker mode is concentrated close to the center plane, displaying spanwise-altering structures (see figure~\ref{fig:resl_wavemaker}(d)).

The wavemaker gains and corresponding spatial structures for the LSL region are displayed in figure~\ref{fig:resl_wavemaker}(e-h).
The wavemaker gains $\zeta$ peak at $St = 0.375$ and 0.4375 for the optimal and first sub-optimal modes, respectively.
For lower frequencies ($St < 0.5$), the optimal wavemaker gain $\zeta_z$ is concentrated near the center plane ($z = 0$).
As the frequency increases ($St > 0.5$), this peak shifts away from the center plane to the spanwise region $-0.3 < z/w_{sp} < -0.2$.
An opposite trend is observed for the first sub-optimal mode, in which the wavemaker gain moves closer to the center plane with increasing frequency, as shown in figure~\ref{fig:resl_wavemaker}(f).
The spatial structure of the optimal wavemaker mode remains close to the center plane, whereas the sub-optimal mode exhibits displacement away from the center (figure~\ref{fig:resl_wavemaker}(g-h)).

\section{Conclusions} \label{sec:concl}

The evolution of spatiotemporal dynamics of a shear layer, comprised of a core and bypass stream, developing in the presence of proximal boundary layers and sidewall-induced corner vortices, is explored using a combination of data-driven modal decomposition method and operator-driven triglobal resolvent analysis.
The spectral proper orthogonal decomposition (SPOD) results highlight two distinct dynamical regimes: tonal and broadband.
The tonal behavior at the dominant frequency ($St = 3.225$) originates from the two-dimensional Kelvin–Helmholtz instability developing in the splitter plate shear layer formed by mixing of core and bypass streams downstream of the splitter plate trailing edge.
Conversely, the upper and lower shear layers, generated by mixing of the bulk jet flow with the ambient, are dominated by broadband, low-frequency ($St \leq 1.5$), and strongly 3D coherent structures.
Bispectral mode decomposition further uncovers nonlinear broadband triadic energy transfer within the upper and lower shear layers at low frequencies, as well as energy transfer from the tonal frequency to its sub- and first harmonics in the splitter plate shear layer.

To investigate the amplification mechanisms driving the coherent structures, triglobal resolvent analysis is carried out in the upper and lower shear layer regions, respectively. In both regions, the resolvent spectra show stronger energy amplification at low frequencies, consistent with the SPOD results. Forcing–response pairs are extracted for both the optimal and sub-optimal modes.
In the initial shear layer, the response is excited by forcing located near the nozzle lip.
The convective speed of the resulting wavepackets is also quantified, showing a constant phase velocity of $c_{ph} = 0.77$ at low frequencies.
These spatial structures and their phase speed highlight the convective nature of the 3D KH instability in the initial regions of the upper and lower shear layers.

The influence of the rectangular jet corner on flow characteristics is elucidated.
The resolvent response modes indicate the formation of relatively low-frequency streamwise vortices originating at the nozzle corner.
The resolvent wavemaker analysis further supports the presence of a self-sustaining vortex generation mechanism in this region at low frequencies, which contributes to axis-switching behavior in rectangular jets.
These findings offer valuable insights into the underlying amplification mechanisms in a multi-stream jet flow and point toward potential flow control strategies not only at suppressing high-frequency resonant tones but also at reducing broadband noise, particularly at low frequencies, in future studies.

\section*{Acknowledgements}
We gratefully acknowledge support from the Air Force Office of Scientific Research (AFOSR) under award FA9550-23-1-0019 (Program Officer: Dr. Gregg Abate). We acknowledge the computational resources provided by Syracuse University Research Computing. This work used Bridges-2 at Pittsburgh Supercomputing Center through allocation PHY250070 from the Advanced Cyberinfrastructure Coordination Ecosystem: Services \& Support (ACCESS) program, which is supported by National Science Foundation grants \#2138259, \#2138286, \#2138307, \#2137603, and \#2138296. The authors would like to thank Dr. Mark N. Glauser and Dr. Fernando Zigunov for fruitful discussions.

\appendix

\section{Validation of 3D linear operator} \label{sec:app_3d_lin_op}

In this section, we first perform the validation of our in-house software to build the 3D linear Navier-Stokes (N-S) operator. The biglobal linear operator framework \citep{sun2017} is extended to a triglobal framework to consider all three inhomogeneous directions in \textit{CharLES} \citep{bres2017}. For validation purposes, a canonical 3D cubic lid-driven cavity flow is considered. The flow conditions are kept the same as described by~\citep{theofilis2017linearized} and~\citep{ohmichi2021matrix}, where the top wall is moving with a specific velocity profile and the other five walls are stationary. The computational domain is set to $[x,y,z] \in [0,0,0] \times [1,1,1]$. The top wall velocity profile is given by
\begin{equation}
    u(x,1,z) = [1-(2x-1)^{16}]^2 [1-(2z-1)^{16}]^2.
\end{equation}

\begin{figure}
    \centering
    \begin{tabular}{c|c|c|c}
    \hline
           & \textbf{1st Mode} & \textbf{2nd Mode} & \textbf{3rd Mode }\\
         \hline \hline
         \cite{theofilis2017linearized}&$-0.4164i + 0.1320$&$-0.4507i + 0$& $-0.5793i + 0.4533$ \\
         \cite{ohmichi2021matrix}      &$-0.4136i + 0.1298$ &$-0.4485i + 0$& $-0.5777i + 0.4471$ \\
         Present Study                 &$-0.4147i + 0.1274$&$-0.4517i + 0$& $-0.5743i + 0.4474 $ \\
    \hline
    \end{tabular}
    \includegraphics[width=1\linewidth]{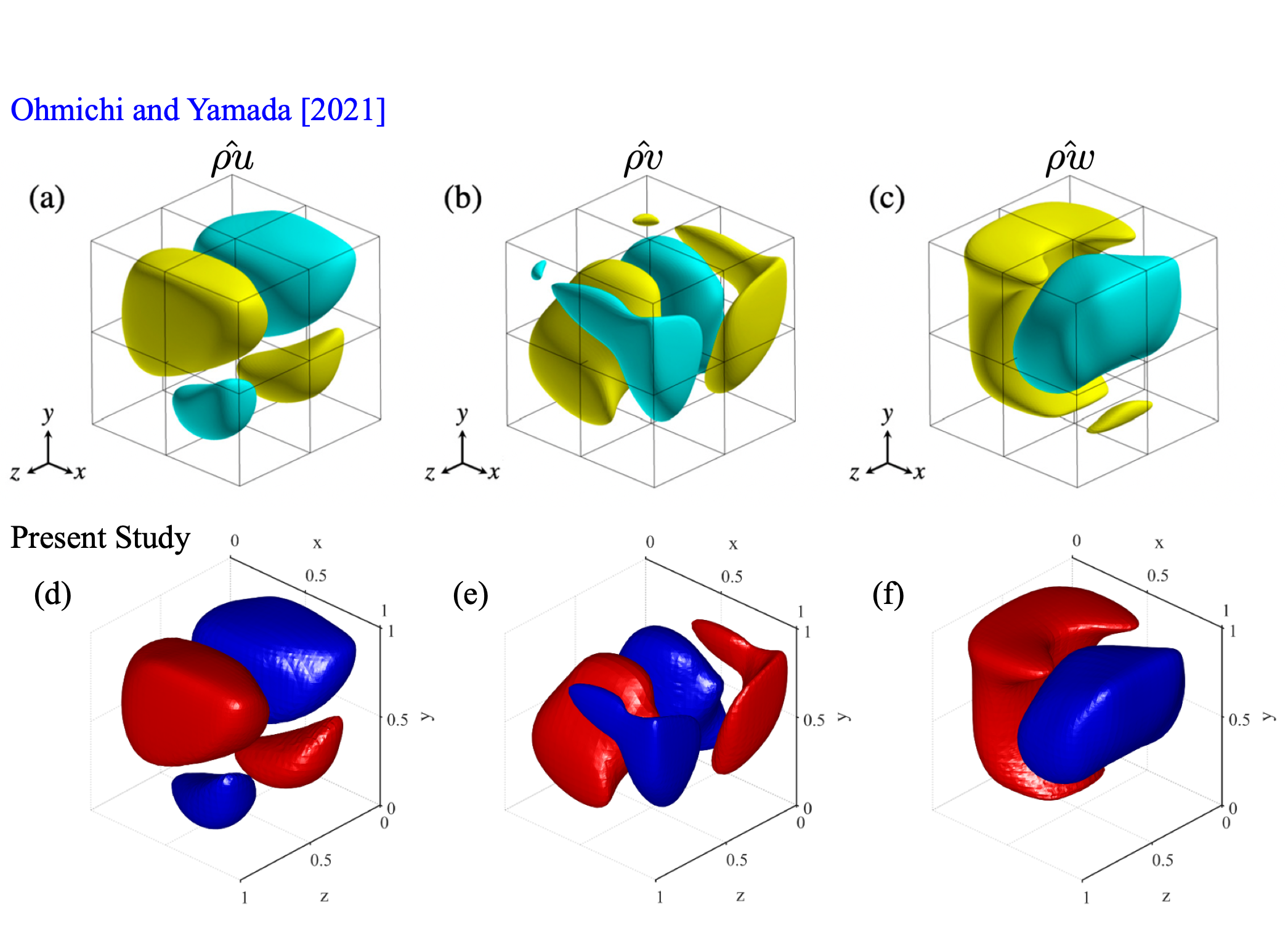}
    \caption{Comparison of eigenvalues for 3D cubic lid-driven cavity flow. The eigenvectors (real component) of $x$-momentum ($\hat{\rho u}$), $y$-momentum ($\hat{\rho v}$) and $z$-momentum ($\hat{\rho w}$) from \cite{ohmichi2021matrix} (a-c) are compared with the present analysis (d-f). \textcolor{blue}{\rule{1.5ex}{1.5ex}}\textcolor{red}{\rule{1.5ex}{1.5ex}} (blue, red) = [-0.002, 0.002].}
    \label{fig:L_vali}
\end{figure}

The time-averaged flow is obtained by using \textit{CharLES} \citep{bres2017} at Mach~$0.2$ and Reynolds number~$200$, based on the velocity of the moving wall.
To obtain a linearized 3D N-S operator, each wall is prescribed a zero Neumann boundary condition for fluctuating pressure and Dirichlet boundary conditions for velocities.
Based on the grid convergence test, the computational domain is discretized with $64^3$ control volumes.
The resulting 3D linearized N-S operator $\boldsymbol{L}(\boldsymbol {\overline{q}})$ is a size of $5 \times 64^3$ in a matrix form.
The eigenvalue decomposition is performed on $\boldsymbol{L}(\boldsymbol {\overline{q}})$ and the first three least stable modes contained by the present analysis are compared in figure~\ref{fig:L_vali}. The eigenvalues are in good agreement with the results obtained by \cite{theofilis2017linearized} and \cite{ohmichi2021matrix}, with a relative difference of $\mathcal{O}(10^{-3})$. The eigenvectors of the momentum from \cite{ohmichi2021matrix} (figure~\ref{fig:L_vali}(a-c)) are compared with the present study (figure~\ref{fig:L_vali}(d-f)) for the least stable mode (1st Mode).
The spatial three-dimensional structures of the instability are similar to those in the previous study, which validates our 3D Navier-Stokes operator.

\section{Randomized algorithm test vector convergence} \label{sec:app_rsvd}

Due to the large and dense nature of the 3D resolvent operator, computing a full singular value decomposition (SVD) is computationally infeasible.
Instead, a randomized SVD, described by~\cite{ribeiro2020}, is utilized to extract the dominant forcing and response mode pairs.
The convergence of the randomized SVD is evaluated by varying the number of test vectors ($k$) and power iterations ($p$) for the USL case at $St = 0.5$.
Table~\ref{table:rsvd_k} summarizes the first five singular values for $k = 5, 10, 20$ and $p = 0, 2$.
Since the analysis primarily focuses on the characteristics of the leading modes, a configuration with $k = 10$ and no power iterations yields sufficiently accurate results, with a difference of less than 3\% in the leading singular values.
\begin{table}
    \centering
    \begin{tabular}{c|c|c|c|c}
    \hline \hline
         $\sigma_i$ & $k = 5, p =0 $& $k = 10, p =0 $ & $k = 10, p =2 $ &  $k = 20, p =0 $\\
         \hline
         $\sigma_1$ &6.23990&6.40075& 6.27580& 6.44950 \\
         $\sigma_2$ &4.71171 &4.84999& 4.82475&4.90694 \\
         $\sigma_3$ & 3.54226&4.39029& 4.26984 &4.46413\\
         $\sigma_4$ &3.27636&3.69475& 3.80097 & 3.83992\\
         $\sigma_5$ &2.53038&2.83324& 3.24969 &  3.26683\\
    \hline \hline
    \end{tabular}
    \caption{First five singular values obtained from randomized singular value decomposition with test vectors $k=5,10,20$ and power iteration $p=0,2$ for the USL case at $St=0.5$.}
    \label{table:rsvd_k}
\end{table}

\bibliographystyle{unsrt}  
\bibliography{references}

\end{document}